\documentclass[twocolumn,showpacs,aps,prd,floatfix,preprintnumbers]{revtex4}
\usepackage{graphicx} 
\usepackage{dcolumn}
\usepackage{epsfig}    
\usepackage{amsmath}
\input babarsym   

\newcommand{\MM}{\ensuremath{M^2_{\mathrm{rec}}}\xspace}

\def\gevcccc{\mbox{${\mathrm{GeV^2}}/c^4\ $}}
\newcommand{\pipiz}{\ensuremath{\pip \pim \piz}\xspace}
\newcommand{\kkpiz}{\ensuremath{\Kp\Km \piz}\xspace}
\newcommand{\kskpi}{\ensuremath{\KS\Kpm \pimp}\xspace}
\newcommand{\psipipiz}{\ensuremath{\jpsi \ \to \ \pip\pim \piz}\xspace}
\newcommand{\psikkpiz}{\ensuremath{\jpsi \ \to \ \Kp\Km\piz}\xspace}
\newcommand{\psikskpi}{\ensuremath{\jpsi \ \to \ \KS\Kpm\pimp}\xspace}
\def\Kstarz   {\ensuremath{K^*_0(1430)}\xspace}

\def\calR         {{\ensuremath{\cal R}\xspace}}
\newcommand{\Real}{\ensuremath{\mathcal{R}e}\xspace}
\newcommand{\BaBarPubYear}    {16}
\newcommand{\BaBarPubNumber}  {008}
\newcommand{\SLACPubNumber}   {16920}
\newcommand{\al}{\ensuremath{\kern 0.5em }}
\newcommand{\all}{\ensuremath{\kern 0.25em }}
\begin{document}
\begin{flushleft}
\babar-PUB-\BaBarPubYear/\BaBarPubNumber \\
SLAC-PUB-\SLACPubNumber \\
\end{flushleft}
\title{
 \large \bf\boldmath Dalitz plot analyses of \psipipiz, \ \psikkpiz, \ and \psikskpi produced via $\epem$ annihilation with initial-state radiation

}
%
\author{J.~P.~Lees}
\author{V.~Poireau}
\author{V.~Tisserand}
\affiliation{Laboratoire d'Annecy-le-Vieux de Physique des Particules (LAPP), Universit\'e de Savoie, CNRS/IN2P3,  F-74941 Annecy-Le-Vieux, France}
\author{E.~Grauges}
\affiliation{Universitat de Barcelona, Facultat de Fisica, Departament ECM, E-08028 Barcelona, Spain }
\author{A.~Palano}\altaffiliation{Also at: Thomas Jefferson National Accelerator Facility,  Newport News, Virginia 23606, USA}
\affiliation{INFN Sezione di Bari and Dipartimento di Fisica, Universit\`a di Bari, I-70126 Bari, Italy }
\author{G.~Eigen}
\affiliation{University of Bergen, Institute of Physics, N-5007 Bergen, Norway }
\author{D.~N.~Brown}
\author{Yu.~G.~Kolomensky}
\affiliation{Lawrence Berkeley National Laboratory and University of California, Berkeley, California 94720, USA }
\author{M.~Fritsch}
\author{H.~Koch}
\author{T.~Schroeder}
\affiliation{Ruhr Universit\"at Bochum, Institut f\"ur Experimentalphysik 1, D-44780 Bochum, Germany }
% ...
\collaboration{The \babar\ Collaboration}
\noaffiliation

\begin{abstract}
We study the processes $\epem \ \to \ \gamma_{\rm ISR} \jpsi$ where \psipipiz, \  \psikkpiz, \ and  \psikskpi using a data sample
of 519~\invfb\ recorded with the \babar\ detector operating at the SLAC PEP-II
asymmetric-energy \epem\ collider at center-of-mass energies at and near the
$\Upsilon(nS)$ ($n = 2,3,4$) resonances. 
We measure the ratio of branching fractions 
$\calR_1 = \frac{\BR(J/\psi \ \to \ K^+ K^- \pi^0)}{\BR(J/\psi \ \to \ \pi^+ \pi^- \pi^0)}$ and
$\calR_2 = \frac{\BR(J/\psi \ \to \ \KS K^{\pm} \pi^{\mp})}{\BR(J/\psi \ \to \ \pi^+ \pi^- \pi^0)}$.
We perform Dalitz-plot analyses of the three \jpsi decay modes and measure fractions for resonances contributing to the decays. 
We also analyze the \psipipiz decay using the Veneziano model. We observe structures compatible with the presence
of $\rho(1450)$ in all the three \jpsi decay modes and measure the relative branching fraction:
$\calR(\rho(1450)) = \frac{\BR(\rho(1450) \ \to \ K^+ K^-)}{\BR(\rho(1450) \ \to \ \pi^+ \pi^-)} = 0.307 \pm 0.084 ({\rm stat}) \pm 0.082 ({\rm sys})$.
\end{abstract}
\pacs{13.25.Gv, 14.40.Pq, 14.40.Rt}

\maketitle

\section{Introduction}

Charmonium decays, in particular radiative and hadronic decays of the $J/\psi$ meson, have been studied extensively~\cite{Kopke:1988cs,Bai:2003ww}.
One of the motivations for these studies is to search for non-$q \bar q$ mesons such as glueballs or molecular 
states that are predicted by
QCD to populate the low mass region of the hadron mass spectrum~\cite{glue}.

Previous studies of $J/\psi$ decays to $\pip \pim \piz$ show a clear signal of
$\rho(770)$ production~\cite{bes1,bes2}. In addition there is indication of
higher mass resonance production in $\psi(2S)$ decays~\cite{bes2}. This is
not necessarily the case in \jpsi decays, but neither does the $\rho(770)$ contribution saturate the spectrum. Attempts have been made
to describe the \jpsi decay distribution with additional partial waves~\cite{adam0}. It was found that
interference effects are strong and even after adding $\pi \pi$
interactions up to $\approx$ 1.6 \gevcc\ the description remained
quite poor. Continuing to expand the partial wave basis
to cover an even higher mass region would lead to a rather
unconstrained analysis. On the other hand with the amplitudes developed in the Veneziano model, all partial waves are
related to the same Regge trajectory, which gives a very strong constraint on the amplitude analysis~\cite{adam}.

While large samples of $J/\psi$ decays exist, some branching
fractions remain poorly measured. In particular the 
$J/\psi \ \to \ \Kp \Km \piz$ branching fraction has been measured by MarkII~\cite{markii} using only 25 events.

Only a preliminary result exists, to date, on a Dalitz-plot analysis of $J/\psi$ decays to $\pip \pim \piz$~\cite{bill}.
The BES III experiment~\cite{bes3} has performed an angular analysis of  $J/\psi \ \to \ \Kp \Km \piz$.
The analysis requires the presence of a broad $J^{PC}=1^{--}$ state in the $K^+K^-$ threshold region, which is interpreted as
a multiquark state.
However Refs.~\cite{int1,int2} explain it by the interference between the $\rho(1450)$ and $\rho(1700)$.
On the other hand, the decay $\rho(1450) \ \to \ \Kp \Km$ appears as ``not seen'' according to the PDG listing~\cite{PDG}.
No Dalitz-plot analysis has been performed to date on the \psikskpi decay.

We describe herein a study of the \psipipiz, \ \psikkpiz, and \psikskpi decays produced in \epem annihilation via initial-state radiation (ISR), where only resonances with $J^{PC}=1^{--}$ can be produced.

This article is organized as follows. In Sec.\ II, a brief description of the
\babar\ detector is given. Section III is devoted to the event reconstruction and data selection. In Sec.\ IV, we describe the study of efficiency and resolution,
while Sec.\ V is devoted to the measurement of the \jpsi branching fractions. Section VI describes the Dalitz-plot analyses while in Sec.\ VII, we report the measurement of the $\rho(1450)$ branching fraction. Finally we summarize the results in Sec. VIII. 

\section{The \babar\ detector and dataset}

The results presented here are based on data collected
with the \babar\ detector
at the PEP-II asymmetric-energy $e^+e^-$ collider
located at SLAC. The data sample corresponds
to an integrated luminosity of 519~\invfb~\cite{lumy} recorded at
center-of-mass energies at and near the $\Upsilon (nS)$ ($n=2,3,4$)
resonances. 
The \babar\ detector is described in detail elsewhere~\cite{BABARNIM}.
Charged particles are detected, and their
momenta are measured, by means of a five-layer, double-sided microstrip detector,
and a 40-layer drift chamber, both operating  in the 1.5~T magnetic 
field of a superconducting
solenoid. 
Photons are measured and electrons are identified in a CsI(Tl) crystal
electromagnetic calorimeter (EMC). Charged-particle
identification is provided by the specific energy loss in
the tracking devices, and by an internally reflecting ring-imaging
Cherenkov detector. Muons are detected in the
instrumented flux return  of the magnet.
Monte Carlo (MC) simulated events~\cite{geant}, with sample sizes 
more than 10  times larger than the corresponding data samples, are
used to evaluate signal efficiency and to determine background features.

\section{Event Reconstruction and Data Selection}

We study the following reactions
\begin{equation}
  \epem \ \to \ \gamma_{\rm ISR} \ \pipiz,
  \label{eq:pi3}
\end{equation}
\begin{equation}
  \epem \ \to \ \gamma_{\rm ISR} \ \kkpiz,
    \label{eq:kkpi0}
\end{equation}
\begin{equation}
  \epem \ \to \ \gamma_{\rm ISR} \ \kskpi,
    \label{eq:k0skpi}
\end{equation}
where $\gamma_{\rm ISR}$ indicates the ISR photon.

For reactions (\ref{eq:pi3}) and (\ref{eq:kkpi0}), we consider only events for which the number of well-measured charged-particle tracks with
transverse momenta greater than 0.1~\gevc\ is exactly equal to two.
The charged-particle tracks are fitted to a common vertex with the requirements that they
originate from the interaction region and that the $\chi^2$ probability of the vertex fit be greater than 0.1\%. 
We observe prominent \jpsi signals in both reactions and optimize the signal-to-background ratio using the data by 
retaining only selection criteria that do not remove significant \jpsi signal.
We require the energy of the less-energetic photon from \piz\ decays to be
greater than 100~\mev. Each pair of photons is kinematically fitted to a \piz\ 
requiring it to emanate from the primary vertex of the event, and with the diphoton
mass constrained to the nominal $\pi^0$ mass~\cite{PDG}.
Due to the soft-photon background, we do not impose a veto on the presence of additional photons in the final state but
we require exactly one \piz\ candidate in each event.
Particle identification is used in two different ways. For reaction (\ref{eq:pi3}), we require two oppositely charged particles to be loosely identified as pions.
For reaction (\ref{eq:kkpi0}), we loosely identify one kaon and require that neither track be a well-identified pion, electron, or muon.
\begin{figure}[h]
\begin{center}
\includegraphics[width=8cm]{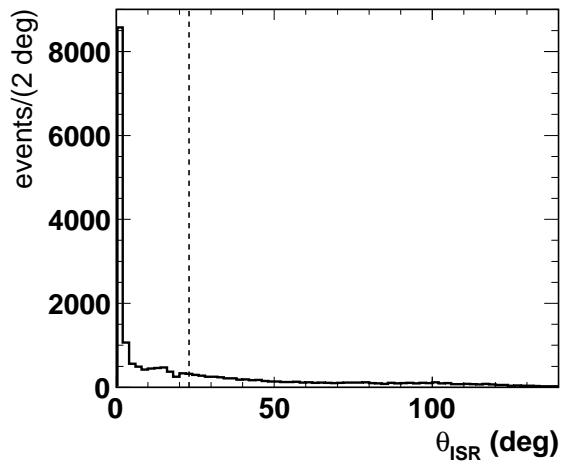}
\caption{(a) Distribution of $\theta_{\rm ISR}$ for events in the $\jpsi \ \to \ \pip \pim \piz$ ISR signal region. The dashed line indicates the $\theta_{\rm ISR}=23^0$ angle.}
\label{fig:fig0}
\end{center}
\end{figure}

\begin{figure}[h]
\begin{center}
  \includegraphics[width=7cm]{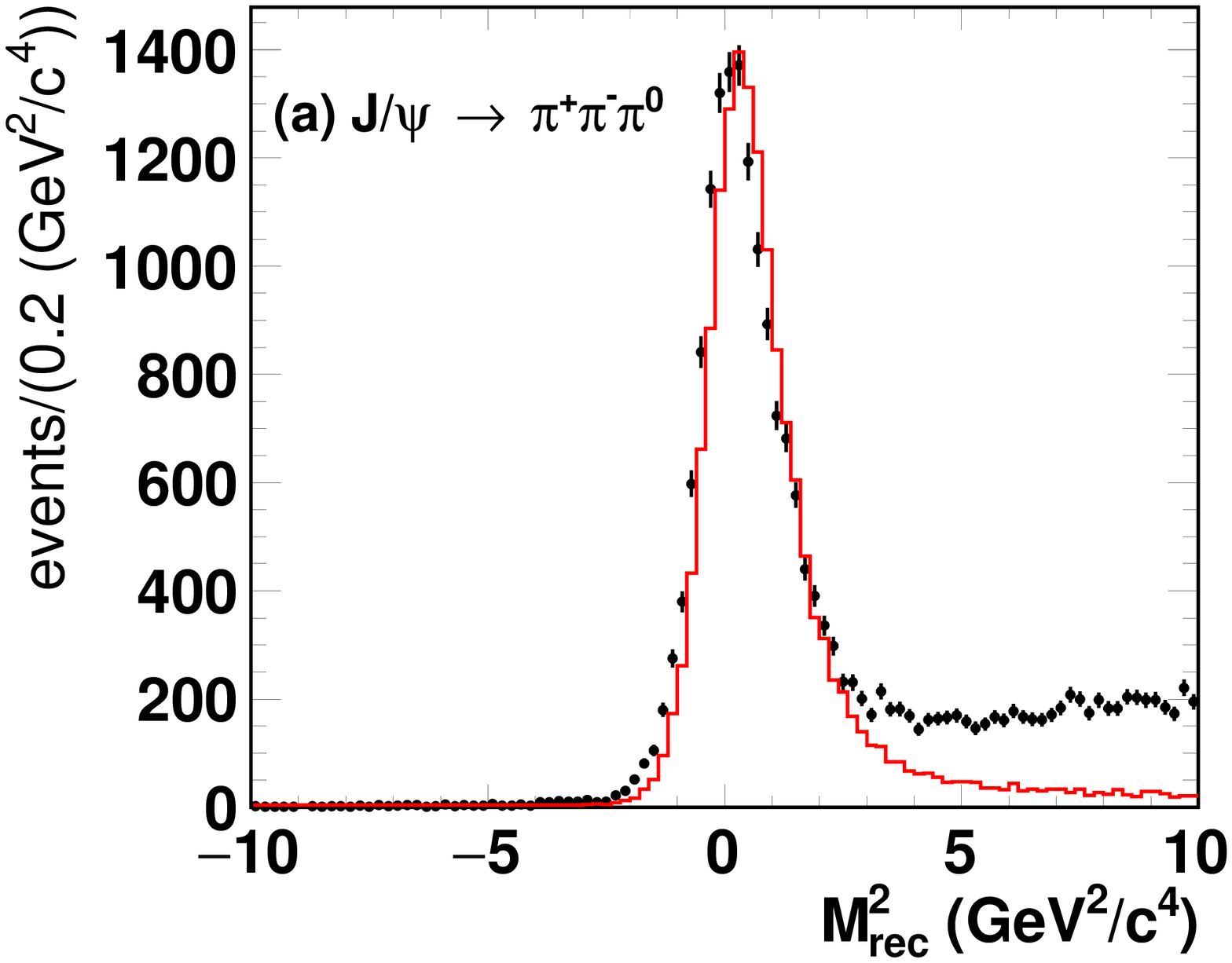}\\
  \includegraphics[width=7cm]{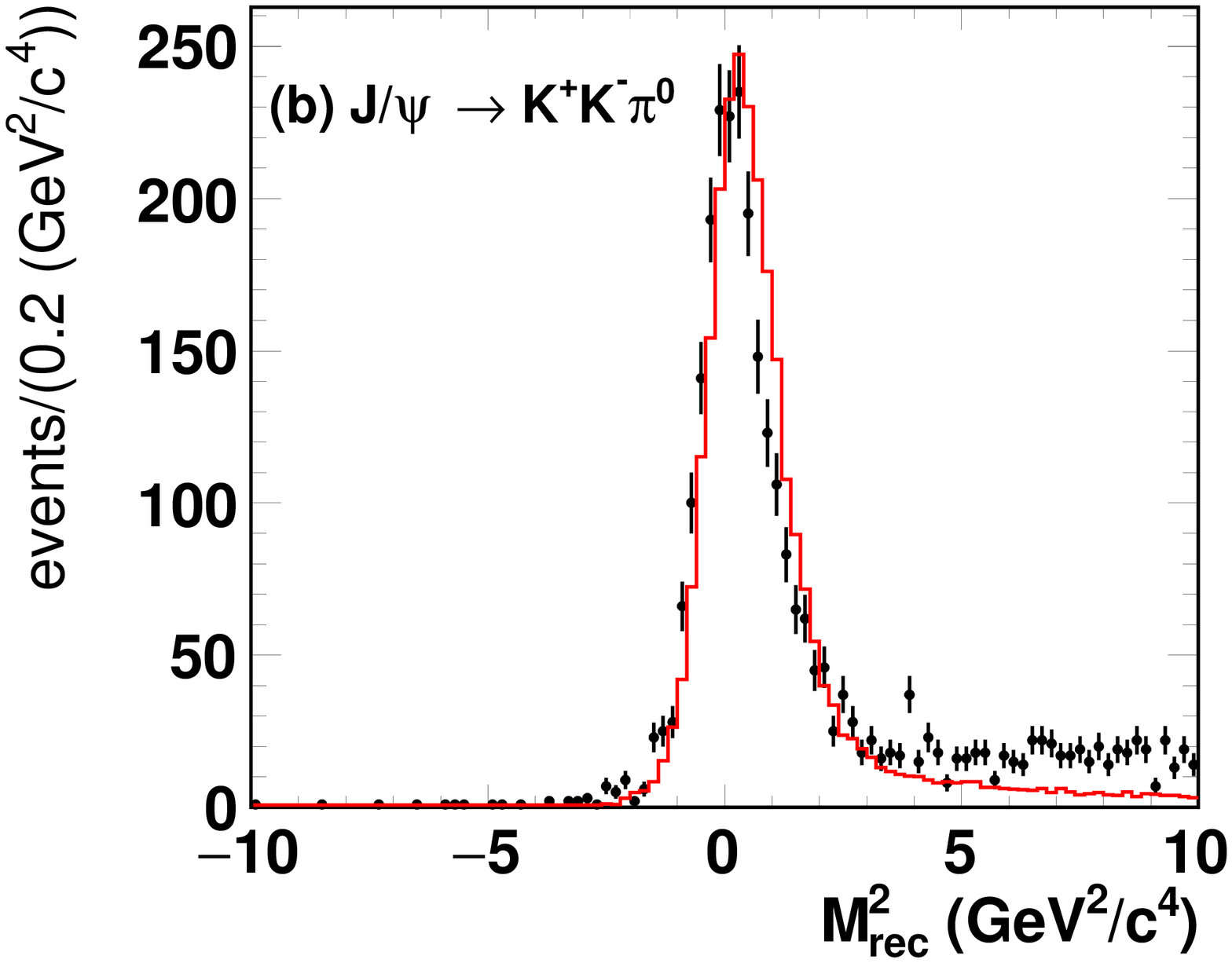}\\
  \includegraphics[width=7cm]{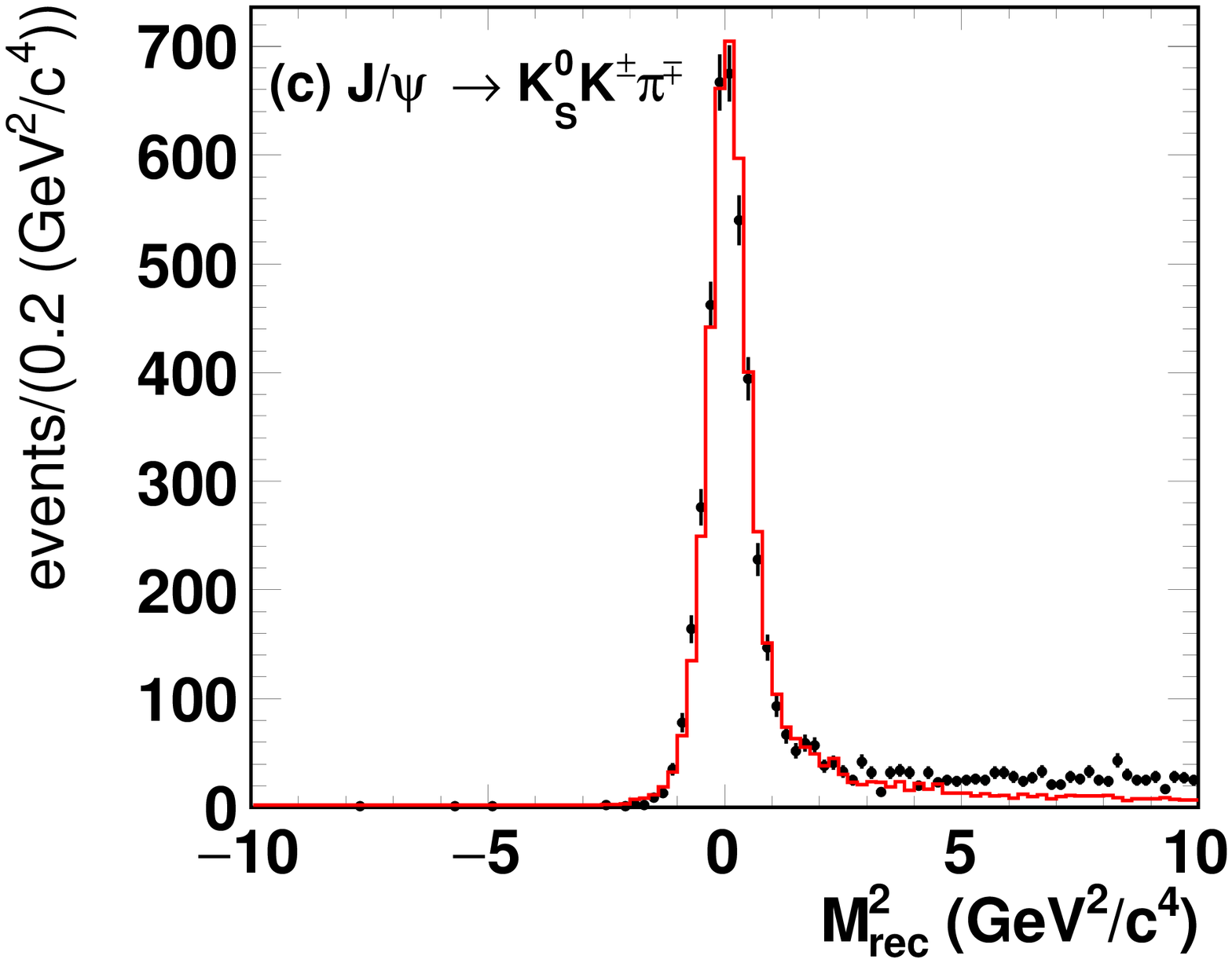}
\caption{Distributions of \MM\ for $\epem \ \to \ \gamma_{\rm ISR} \jpsi$ where (a) $\jpsi \ \to \ \pipiz$,  (b) $\jpsi \ \to \ \kkpiz$, and (c) $\jpsi \ \to \ \kskpi$.
In each figure the data are shown as points with error bars,
and the MC simulation is shown as a histogram.}
\label{fig:fig1}
\end{center}
\end{figure}

\begin{figure}[h]
\begin{center}
  \includegraphics[width=7.0cm]{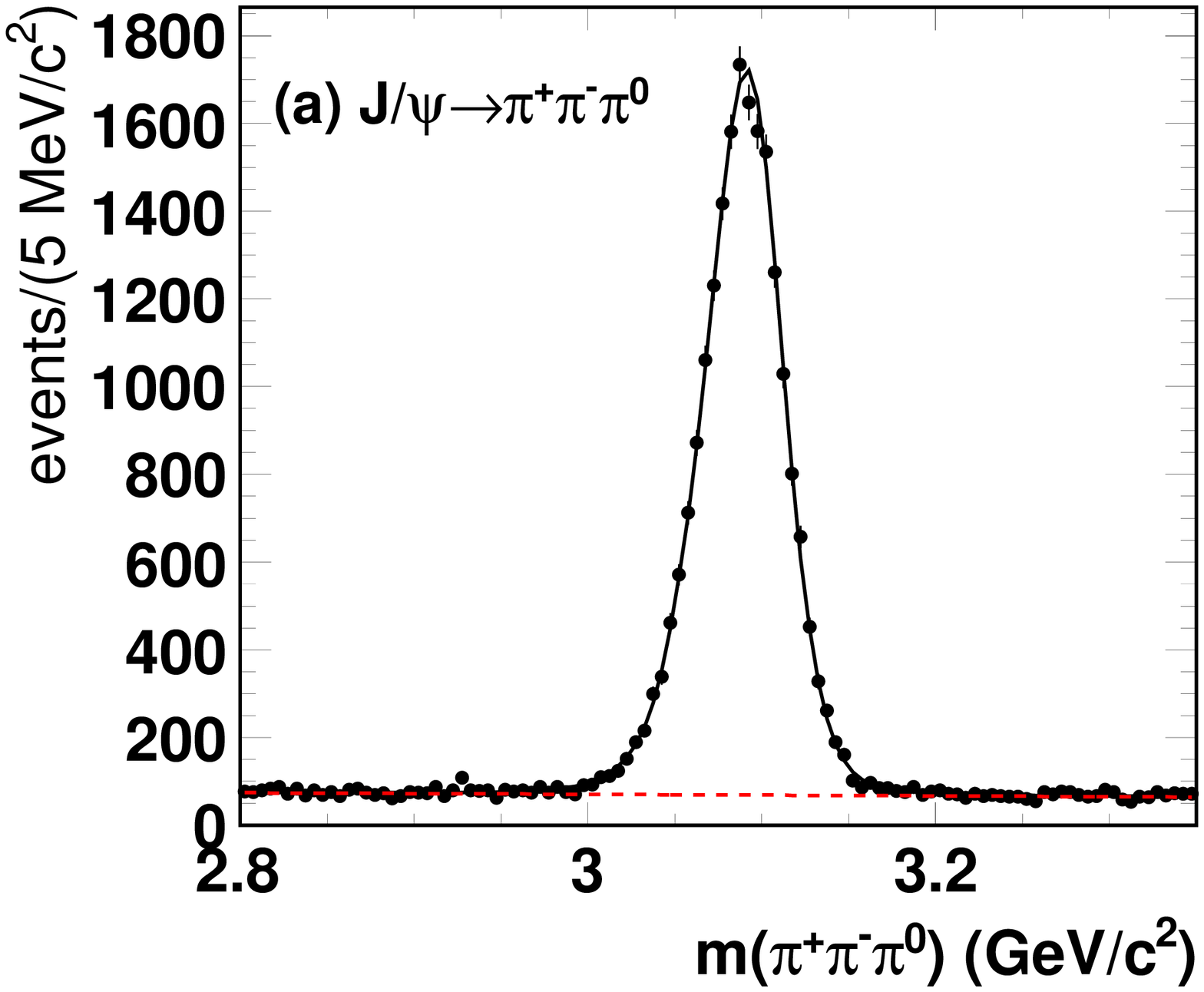}
  \includegraphics[width=7.0cm]{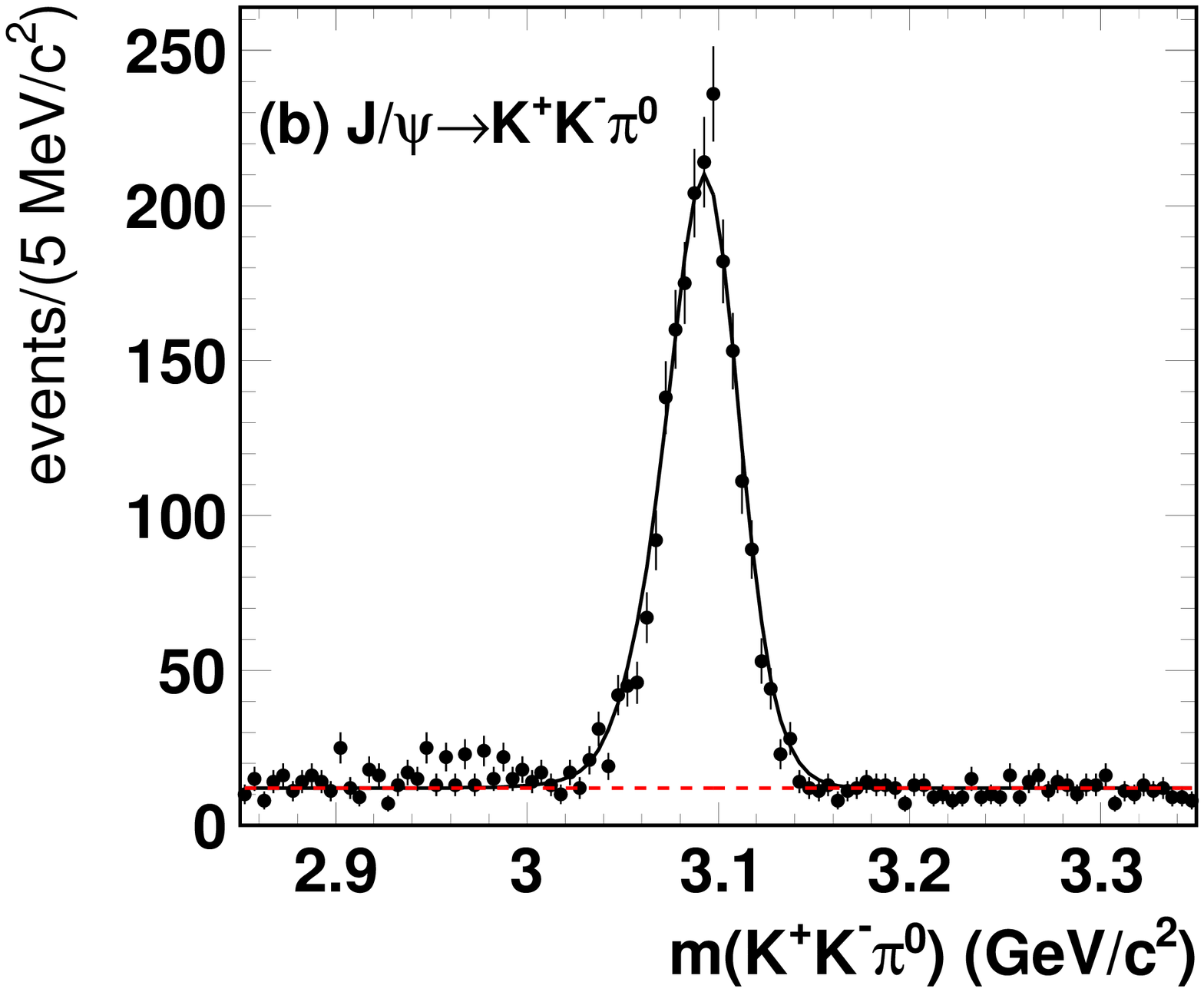}
  \includegraphics[width=7.0cm]{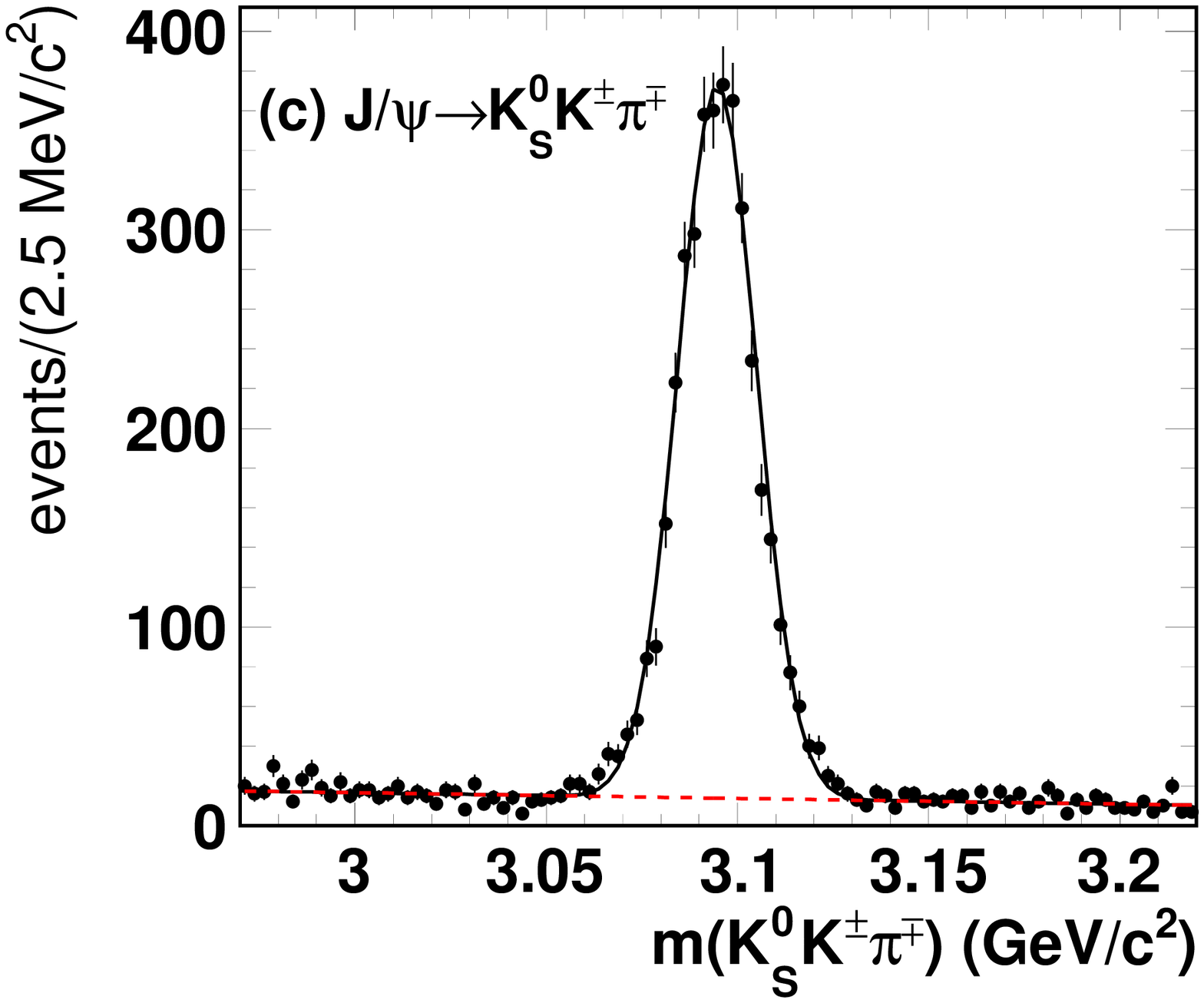}
\caption{(a) The \pipiz\, (b) \kkpiz, and \kskpi mass spectra in the ISR region. In each figure, the solid curve shows the total fit function
and the dashed curve shows the fitted background contribution.}
\label{fig:fig2}
\end{center}
\end{figure}

For reaction (\ref{eq:k0skpi}), we consider only events for which the number of well-measured charged-particle tracks with
transverse momentum greater than 0.1~\gevc\ is exactly equal to four, and for which there are no more than five photon candidates
with reconstructed energy in the EMC greater than 100~\mev.
We obtain $\KS \to \pip \pim$ candidates by means of a vertex fit of pairs of oppositely charged tracks, for which we require a $\chi^2$ fit probability 
greater than 0.1\%. Each \KS candidate is then combined with 
two oppositely charged tracks, and fitted to a common vertex, with the requirements that the fitted vertex be within the
$e^+ e^-$ interaction region and have a $\chi^2$ fit probability greater than 0.1\%.
We select kaons and pions by applying high-efficiency particle identification criteria.
We do not apply any particle identification requirements
to the pions from the \KS decay.
We accept only \KS candidates with decay lengths from the \jpsi candidate decay vertex greater than 0.2 cm, and
require $\cos \theta_{\KS}>0.98$, where $\theta_{\KS}$ is defined as the angle between the \KS momentum direction and the
line joining the \jpsi and \KS vertices.
A fit to the $\pip \pim$ mass spectrum using a linear function for the background and a Gaussian
function with mean $m$ and width $\sigma$ gives $m=497.24$ \mevcc\ and $\sigma=2.9$ \mevcc. We select the $\KS$ signal region to be within
$\pm 2 \sigma$ of $m$ and reconstruct the \KS 4-vector by summing the three-momenta of the pions and computing the energy using the known \KS mass~\cite{PDG}.

The ISR photon is preferentially emitted at small angles 
with respect to the beam axis (see Fig.~\ref{fig:fig0}), and escapes detection in the majority of ISR
events. Consequently, the ISR photon is treated as a missing particle.

We define the squared mass \MM\ recoiling against the \pipiz, \kkpiz, and \kskpi systems using the four-momenta of the beam
particles ($p_{e^\pm}$) and of the reconstructed final state particles:
\begin{equation}
\MM \equiv ( p_{e^-}+p_{e^+}-p_{h_1}-p_{h_2}-p_{h_3})^2,
\end{equation}
where the $h_i$ indicate the three hadrons in the final states.
This quantity 
should peak near zero for both ISR events and for exclusive production of 
$e^+ e^- \ \to \ h_1 h_2 h_3$. However, in the exclusive production
the $h_1 h_2 h_3$ mass distribution peaks at the kinematic limit.
We select the ISR reactions (in the following also defined as ISR regions) requiring
\begin{equation}
|\MM| <2 \ \gevcccc
\end{equation}
for reaction (\ref{eq:pi3}) and (\ref{eq:kkpi0}) and
\begin{equation}
|\MM| <1.5 \ \gevcccc
\end{equation}
for reaction (\ref{eq:k0skpi}). 

We reconstruct the three-momentum of the ISR photon from momentum conservation as
\begin{equation}
  {\bf p}_{ISR} = {\bf p}_{e^-}+{\bf p}_{e^+}-{\bf p}_{h_1}-{\bf p}_{h_2}-{\bf p}_{h_3}.
\end{equation}
Table~\ref{tab:tab0} gives the ranges used to define the ISR signal regions for the three \jpsi decay modes.
\begin{table}[htb]
\caption{Ranges used to define the \jpsi signal regions, event yields, and purities for the three \jpsi decay modes.}
\label{tab:tab0}
\begin{center}
\begin{tabular}{lcrc}
\hline
$J/\psi$  & Signal region  & Event  & Purity \cr
decay mode & (\gevcc) & yields & \% \cr
\hline
  \noalign{\vskip2pt}
$\pip \pim \piz$ & 3.028-3.149 & 20417 & 91.3 $\pm$ 0.2 \cr
$\Kp \Km \piz$ & 3.043-3.138 & 2102 & 88.8 $\pm$ 0.7 \cr
$\KS \Kpm \pimp$ & 3.069-3.121& 3907 & 93.1 $\pm$ 0.4 \cr
\hline
\end{tabular}
\end{center}
\end{table}
We show in Fig.~\ref{fig:fig0}, for events in the $\jpsi \ \to \ \pip \pim \piz$ ISR signal region, the distribution of $\theta_{\rm ISR}$, the angle of the reconstructed ISR photon with respect to the \en beam direction in the laboratory system. We observe a narrow peak close to zero with a tail extending up to $140^0$ while background events from \jpsi sidebands are distributed over the full angular range. Since angular coverage of the EMC starts at $\theta>23^0$,
we improve the signal to background ratio for $J/\psi$ events where $\theta_{\rm ISR}>23^0$, by removing events for which no photon shower is found in the EMC in the expected
angular region. Therefore, we require the difference between the predicted polar and azimuthal angles from $p_{ISR}$ and the closest photon shower to be $|\Delta \theta|<0.1 \rad$ and $|\Delta \phi|<0.05 \rad$. We do not use the information on the energy since some photons may not be fully contained in the EMC.

For reaction (\ref{eq:pi3}) we define the helicity angle $\theta_h$ as the angle in the $\pi^+\pi^-$ 
rest frame between the direction of the $\pi^+$ and the boost from the $\pip \pim$.
We observe that residual background from $\epem \ \to \ \gamma \pip \pim$ is concentrated at $| \cos \theta_{\pi} | \approx 1$ and therefore
we remove events having $| \cos \theta_{\pi} | > 0.95$.
A very small \jpsi signal is observed in the events removed by this selection.
No evidence is found for background from the ISR reaction $\epem \ \to \ \gamma_{\rm ISR}  \kp \km$.

Figure~\ref{fig:fig1} shows the \MM\ distributions for the three reactions in the \jpsi signal regions, in comparison to the corresponding \MM\ distributions obtained from simulation. A peak at zero is observed in all distributions indicating
the presence of the ISR process. 
We observe some discrepancy for reactions (\ref{eq:pi3}) and (\ref{eq:kkpi0}) due to some inaccuracy in reconstructing slow \piz\ in the EMC.
Figure~\ref{fig:fig2} shows the \pipiz, \kkpiz, and \kskpi  mass spectra in the ISR region, before applying the efficiency correction. 
We observe strong \jpsi signals over relatively small backgrounds and no more than one candidate per event.
We perform a fit to the \pipiz, \kkpiz and \kskpi mass spectra. Backgrounds are described by first-order polynomials, and each resonance is represented by a simple Breit-Wigner function convolved with the corresponding resolution function (see Sect. IV). Figure~\ref{fig:fig2} shows the fit result, and Table~\ref{tab:tab1} summarizes the mass values and yields. We observe (not taking into account systematic uncertainties) a $J/\psi$ mass shift of +2.9, +4.1, and -2.2 \mevcc\
for the three decay modes.

\begin{table*}[htb]
\caption{Results from the fits to the mass spectra and efficiency corrections. Errors are statistical only.}
\label{tab:tab1}
\begin{center}
\begin{tabular}{ccccc}
\hline
$J/\psi$ decay mode & $\chi^2/NDF$ & \jpsi mass (\mevcc) & Signal yield & $1/\epsilon$ \cr
\hline
  \noalign{\vskip2pt}
$\pip \pim \piz$ & 90/105 & 3099.8 $\pm$ 0.2 & 19560 $\pm$ 164 & 15.57  $\pm$ 1.05 \cr
$\Kp \Km \piz$ & 129/95 & 3101.0 $\pm$ 0.2 & 2002 $\pm$ 48 & 18.31 $\pm$ 0.63 \cr
$\KS \Kpm \pimp$ & 127/96 & 3094.7 $\pm$ 0.2 & 3694 $\pm$ 64 & 15.15 $\pm$ 0.33 \cr
\hline
\end{tabular}
\end{center}
\end{table*}

\begin{figure}[h]
\begin{center}
  \includegraphics[width=6.5cm]{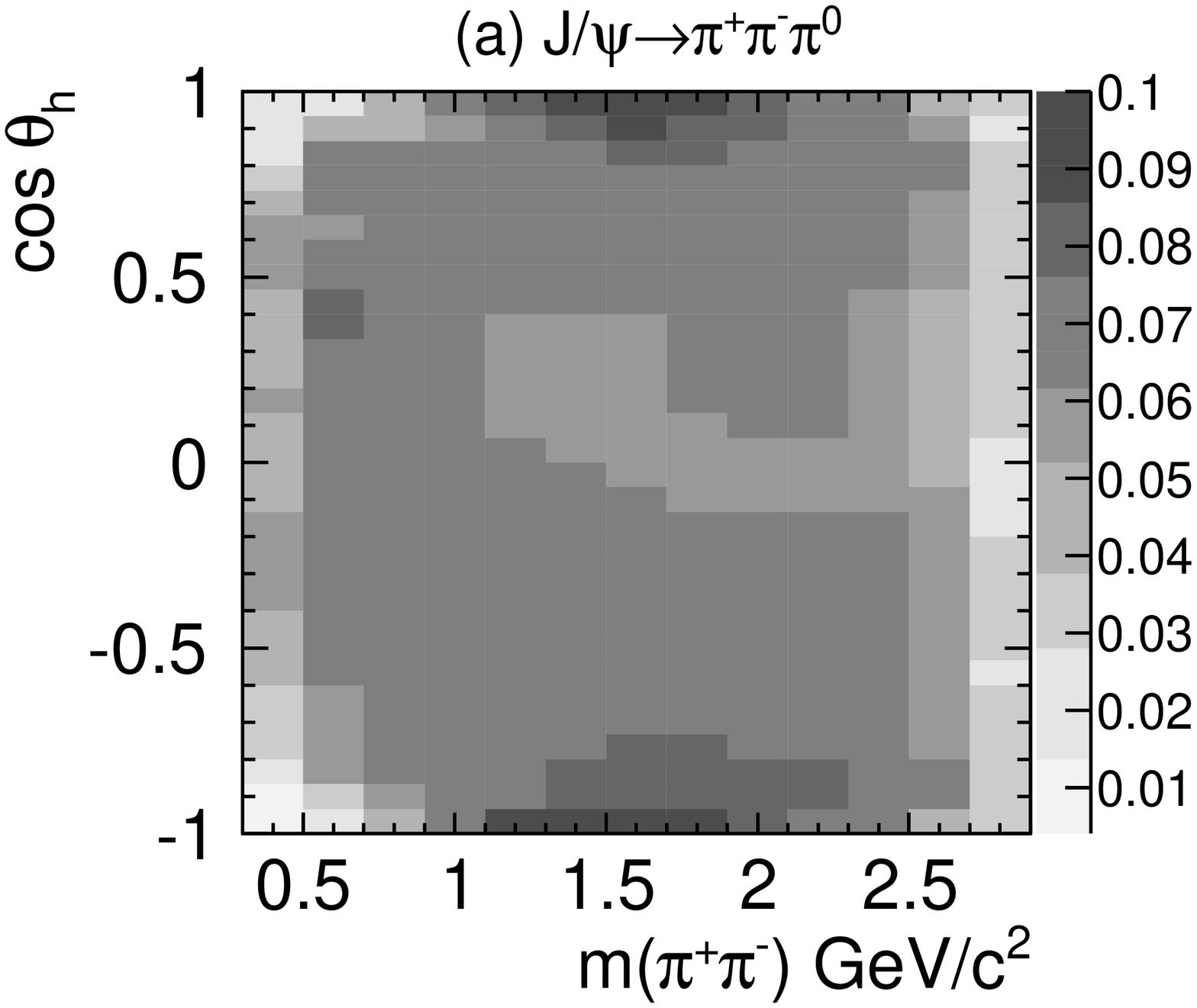}\\
  \includegraphics[width=6.5cm]{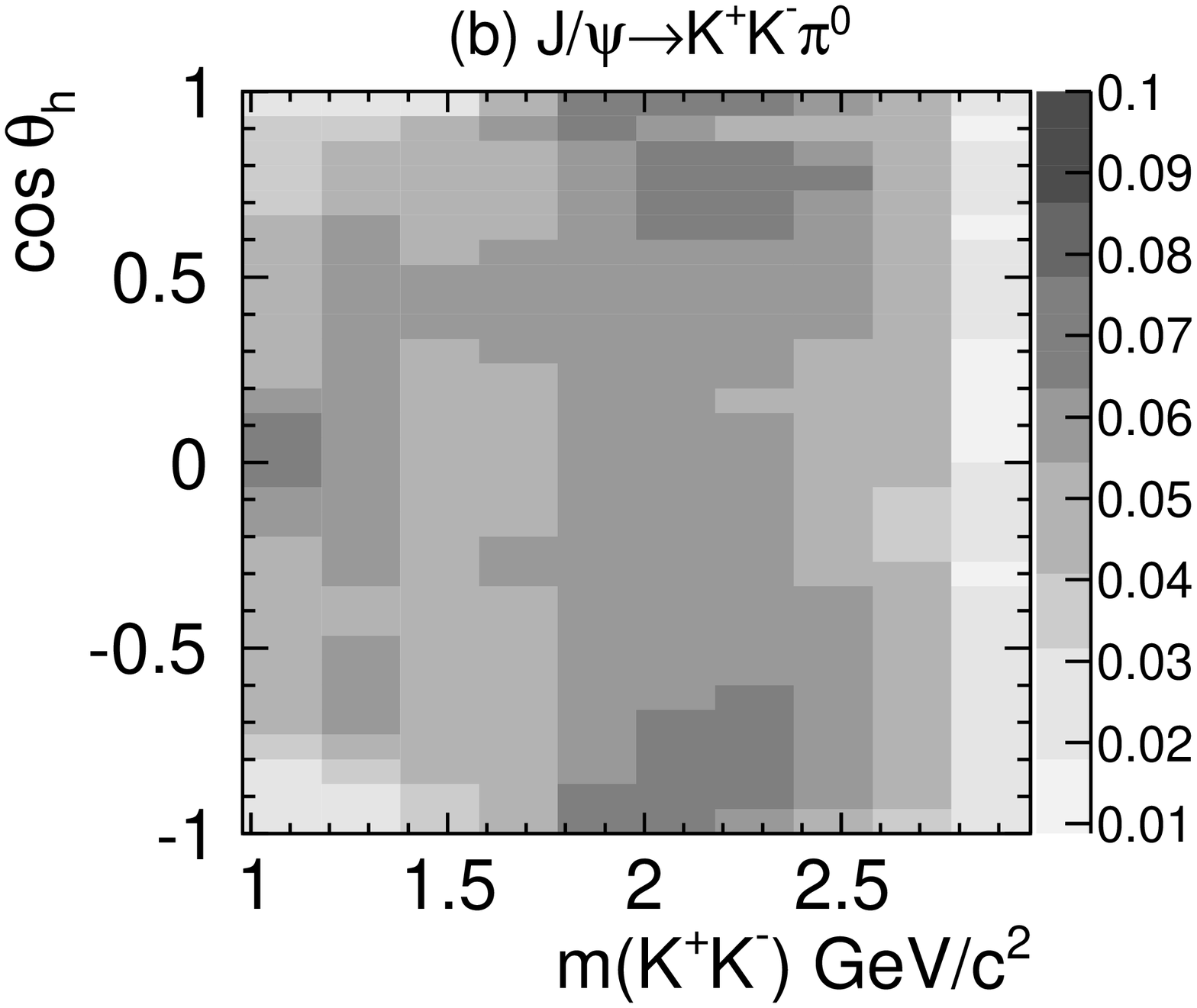}\\
  \includegraphics[width=6.5cm]{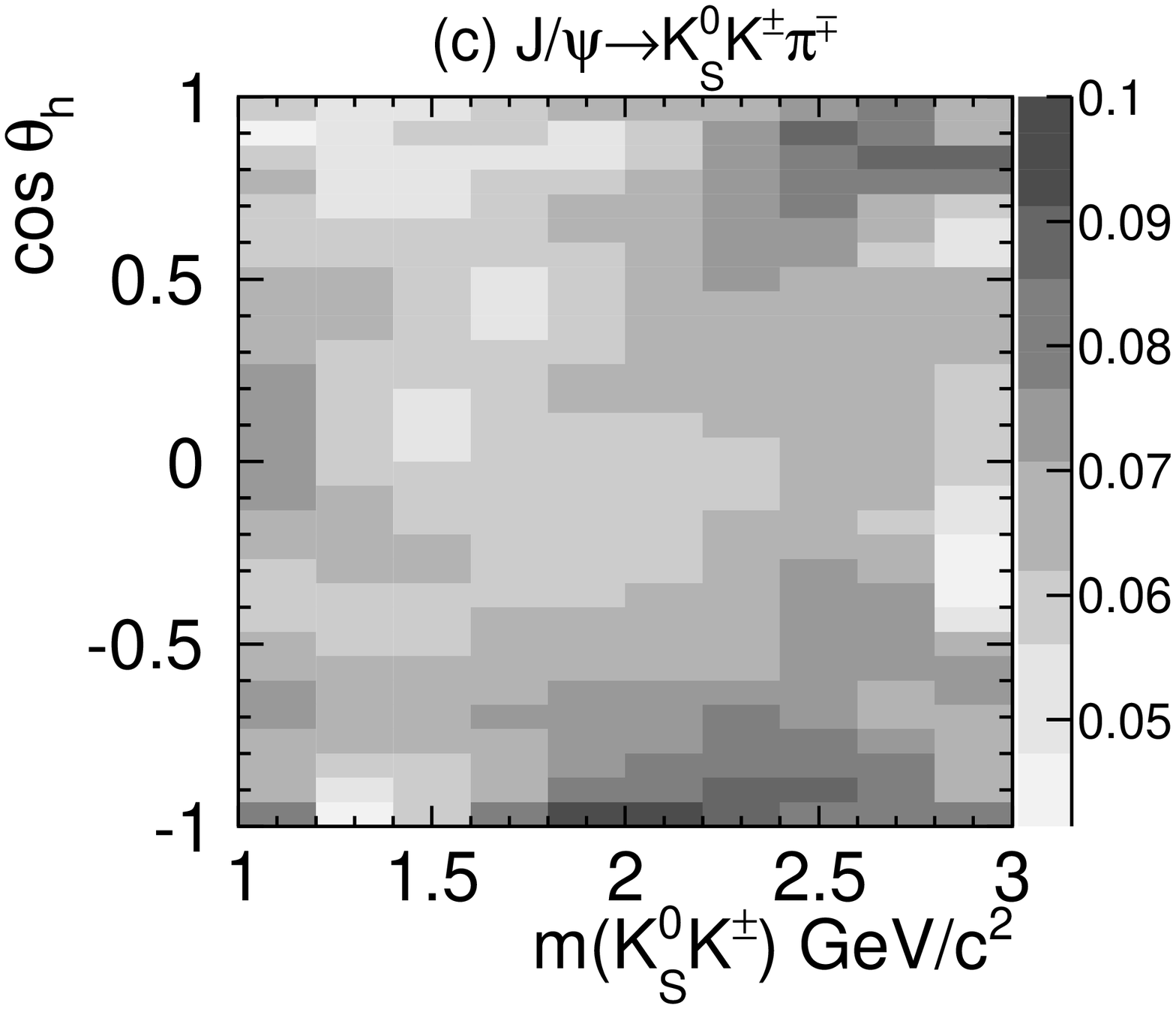}
\caption{Fitted detection efficiency in the $\cos \theta_h \  vs. \ m_{12}$ plane for (a) $\jpsi \ \to \ \pipiz$,   
(b) $\jpsi \ \to \ \kkpiz$, and (c) $\jpsi \ \to \ \kskpi$. Each bin shows the average value of the fit in that region.}
\label{fig:fig3}
\end{center}
\end{figure}

\section{Efficiency and resolution}
\label{eff}

To compute the efficiency, \jpsi MC signal events for the three channels are generated using a detailed detector simulation~\cite{geant} in which the \jpsi decays uniformly in phase space.
These simulated events are reconstructed and analyzed in the same manner as data. The efficiency is computed as the ratio of reconstructed to
generated events.  
We express the efficiency as a function of the $m_{12}$ mass ($\pip \pim$ for $\jpsi \ \to \ \pipiz$,
$\Kp \Km$ for $\jpsi \ \to \ \kkpiz$, and $\KS \Kpm$ for $\jpsi \ \to \ \kskpi$) and $\cos \theta_h$ defined in Sec. III. 
To smooth statistical fluctuations, this efficiency is then parameterized as follows~\cite{babar_z}.

First we fit the efficiency as a function of  $\cos \theta_h$ in separate intervals of $m_{12}$, in terms 
of Legendre polynomials up to $L=12$:
\begin{eqnarray}
\epsilon(\cos\theta_h) = \sum_{L=0}^{12} a_L(m_{12}) Y^0_L(\cos\theta_h).
\end{eqnarray}
For each value of $L$, we fit the mass dependent coefficients $a_L(m_{12})$ with a seventh-order polynomial in $m_{12}$.
Figure~\ref{fig:fig3} shows the resulting fitted efficiency $\epsilon(m_{12},\cos \theta_h)$ for each of the three reactions. 
We observe a significant decrease in
efficiency at low $m_{12}$ for $\cos\theta \sim \pm 1$ and $1.1<m(\Kp \Km)<1.5~\gevcc$ due to the difficulty of reconstructing low-momentum tracks ($p<200$ \mevc in the laboratory frame), which arise because of significant energy losses in the beampipe and inner-detector material.

The mass resolution, $\Delta m$, is measured as the difference between the generated and reconstructed \pipiz\, \kkpiz\, and \kskpi invariant-mass values.
These distributions, for the $J/\psi$ decays having a \piz\ in the final state, deviate from Gaussian
shapes due to a low-energy tail caused by the response of the CsI calorimeter to photons. We fit the distributions using the
sum of a Crystal Ball function~\cite{cb} and a Gaussian function. 
The root-mean-squared values are 24.4  and 22.7 \mevcc \  for the  $\jpsi \ \to \ \pipiz$ and  $\jpsi \ \to \ \kkpiz$ final states, respectively.
The mass resolution for \psikskpi is well described by a single Gaussian having a $\sigma=9.7 \ \mevcc$.

\section{{\boldmath$\protect J/\psi$} Branching Ratios}

We compute the ratio of the branching fractions for $\jpsi \ \to \ \kkpiz$ and $\jpsi \ \to \ \pipiz$ according to
\begin{equation}
\begin{split}
\calR_1 = &\frac{\BR(\jpsi \ \to \ K^+ K^- \piz)}{\BR(\jpsi \ \to \ \pip \pim \piz)} \\
  = &\frac{N_{K^+ K^- \pi^0}}{N_{\pi^+ \pi^- \pi^0}}\frac{\epsilon_{\pi^+ \pi^- \pi^0}}{\epsilon_{K^+ K^- \pi^0}},
\end{split}
\label{eq:r1}
\end{equation}
where $N_{\pi^+ \pi^- \pi^0}$ and $N_{K^+ K^- \pi^0}$ represent the fitted yields for \jpsi in the \pipiz and \kkpiz mass spectra, while $\epsilon_{\pi^+ \pi^- \pi^0}$ and $\epsilon_{K^+ K^- \pi^0}$ are the corresponding efficiencies. 
We estimate $\epsilon_{\pi^+ \pi^- \pi^0}$ and $\epsilon_{K^+ K^- \pi^0}$ for the \jpsi signals by making use of the \mbox{2-D} efficiency 
distributions described in Sec. IV.
To remove the dependence of the fit quality on the efficiency functions we make use of the unfitted efficiency distributions.
Due to the presence of non-negligible backgrounds in the \jpsi signals, which have different distributions in the
Dalitz plot, we perform a sideband subtraction by assigning a weight $w=f/\epsilon(m_{12},\cos \theta)$, where $f=1$ for events in 
the \jpsi signal region and $f=-1$ for events in the sideband regions. The size of the sum of the two sidebands is taken to be the same as that of the signal region.
Therefore we obtain the weighted efficiencies as

\begin{equation}
\epsilon_{h^+ h^- \pi^0} = \frac{\sum_{i=1}^{N}f_i}{\sum_{i=1}^{N}f_i/\epsilon(m_{12},\cos \theta_i)},
\end{equation}
where $N$ indicates the number of events in the signal+sidebands regions.
The resulting yields and efficiencies are reported in Table~\ref{tab:tab1}.

We note that in Eq.~(\ref{eq:r1}) the number of charged-particle tracks
and $\gamma$'s is the same in the numerator and in the denominator of the ratio, so that several systematic uncertainties cancel.
We estimate the systematic uncertainties as follows.
We modify the signal fitting function, describing the \jpsi signals using the sum of two Gaussian functions.
The uncertainty due to efficiency weighting
is evaluated by computing 1000 new weights obtained by randomly modifying the weight in each cell of the 
$\epsilon(m_{12},\cos \theta)$ plane according to its
statistical uncertainty. 
The widths of the resulting Gaussian distributions yield the estimate of the systematic uncertainty for the efficiency weighting procedure. These values are reported as the uncertainties on $1/\epsilon$ in Table~\ref{tab:tab1}.
We assign a 1\% systematic uncertainty for the identification of each of the two kaons, from studies performed using high statistics control samples.
The contributions to the systematic uncertainties from different sources are given in Table~\ref{tab:tab2} and combined in quadrature.
We obtain:
\begin{equation}
\begin{split}
\calR_1 = &\frac{\BR(\jpsi \ \to \ \Kp \Km \piz)}{\BR(\jpsi \ \to \ \pip \pim \piz)} \\
        = &0.120 \pm 0.003 ({\rm stat}) \pm 0.009 ({\rm sys}).
    \label{eq:rr1}
\end{split}
\end{equation}
The PDG reports $\BR(\jpsi \ \to \ \pip \pim \piz)=(2.11 \pm 0.07) \times 10^{-2}$, while
the branching fraction $\BR(\jpsi \to \Kp \Km \piz)$ has been measured by Mark II~\cite{markii} using 25 events, to be $(2.8 \pm 0.8) \times 10^{-3}$.
These values give a ratio $\calR_1^{PDG} = 0.133 \pm 0.038$, in agreement with our measurement.

We perform a test of the $\calR_1$ measurement using a minimum bias procedure. We remove all the selections used to separate reactions (\ref{eq:pi3}) and (\ref{eq:kkpi0}),
except for the requirements on \MM and obtain the events yield for \psipipiz. To obtain the \psikkpiz\ yield, we apply a very loose identifications of the two kaons to remove the large
background and the strong cross-feed from the \psipipiz\ final state. We observe a loss of the $J/\psi$ signal which is estimated by MC to be 3.6\%.
The ratios between the two minimum bias yields, corrected for the above efficiency loss gives directly the ratio of the two branching fractions which is in good agreement with the previous
estimate.

Using a similar procedure as for the measurement of $\calR_1$, correcting for unseen \KS decay modes, we compute the ratio of the branching fractions for $\jpsi \ \to \ \kskpi$ and $\jpsi \ \to \ \pipiz$ according to
\begin{equation}
\begin{split}
\calR_2 = &\frac{\BR(\jpsi \to \KS K^\pm \pi^\mp)}{\BR(\jpsi  \to  \pip \pim \piz)} \\
  = &\frac{N_{\KS K^\pm \pi^\mp}}{N_{\pi^+ \pi^- \pi^0}}\frac{\epsilon_{\pi^+ \pi^- \pi^0}}{\epsilon_{\KS K^\pm \pi^\mp}} \\
  = & 0.265 \pm 0.005 ({\rm stat}) \pm 0.021 ({\rm sys}).
\end{split}
    \label{eq:r2}
\end{equation}

Systematic uncertainties on the evaluation of $\calR_2$ include 0.46\% per track for charged tracks reconstruction, 3\% and 1.1\% for \piz \ and \KS reconstruction, 0.5\% and 1\% for the identification of pions and kaons, respectively. The contributions to the total systematic uncertainty are summarized in Table~\ref{tab:tab2}.

The branching fraction $\BR(\jpsi \to \KS \Kpm \pimp)$ has been measured by Mark I~\cite{marki}, using 126 events, to be $(26 \pm 7) \times 10^{-4}$.
Using the above measurements we obtain an estimate of $\calR_2$:
\begin{equation}
\calR_2^{PDG}=0.123 \pm  0.033,
\end{equation}
which deviates by 3.6$\sigma$ from our measurement.

As a cross check, using the above $\calR_1$ and $\calR_2$ measurements and adding in quadrature statistical and systematic uncertainties, we compute 

\begin{equation}
\calR_3 = \frac{\BR(\jpsi \to \KS K^\pm \pi^\mp)}{\BR(\jpsi  \to  \Kp \Km \piz)} = 2.21 \pm 0.24
\end{equation}

in agreement with the expected value of 2.

\begin{table}[htb]
\caption{Fractional systematic uncertainties in the evaluation of the ratios of branching fractions.}
\label{tab:tab2}
\begin{center}
\begin{tabular}{lcc}
\hline
Effect & $\calR_1$ (\%) & $\calR_2$ (\%)\cr
\hline
  \noalign{\vskip2pt}
Efficiency & 7.5 & 7.0 \cr
Background subtraction & 1.3 & 1.0\cr
Particle identification & 2.0 & 1.8 \cr
\KS reconstruction & & 1.1\cr
\piz reconstruction & & 3.0 \cr
Mass fits & 0.8 & 0.8 \cr
\hline
Total & 7.9 & 8.0 \cr
\hline
\end{tabular}
\end{center}
\end{table}
\begin{figure}
\begin{center}
\includegraphics[width=9cm]{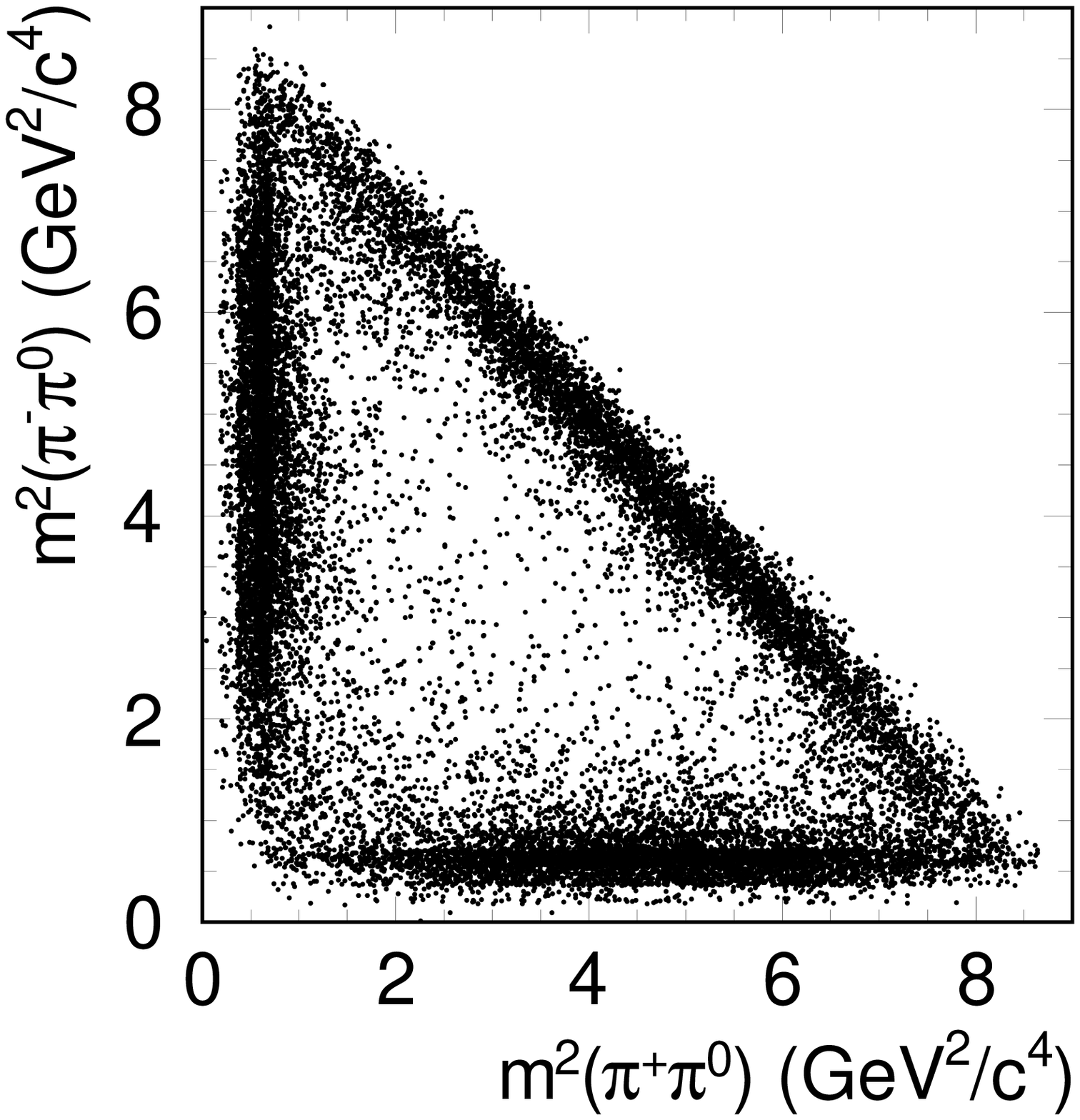}
\caption{Dalitz plot for the \psipipiz\ events in the signal region.}
\label{fig:fig4}
\end{center}
\end{figure}

\begin{figure}
\begin{center}
\includegraphics[width=9cm]{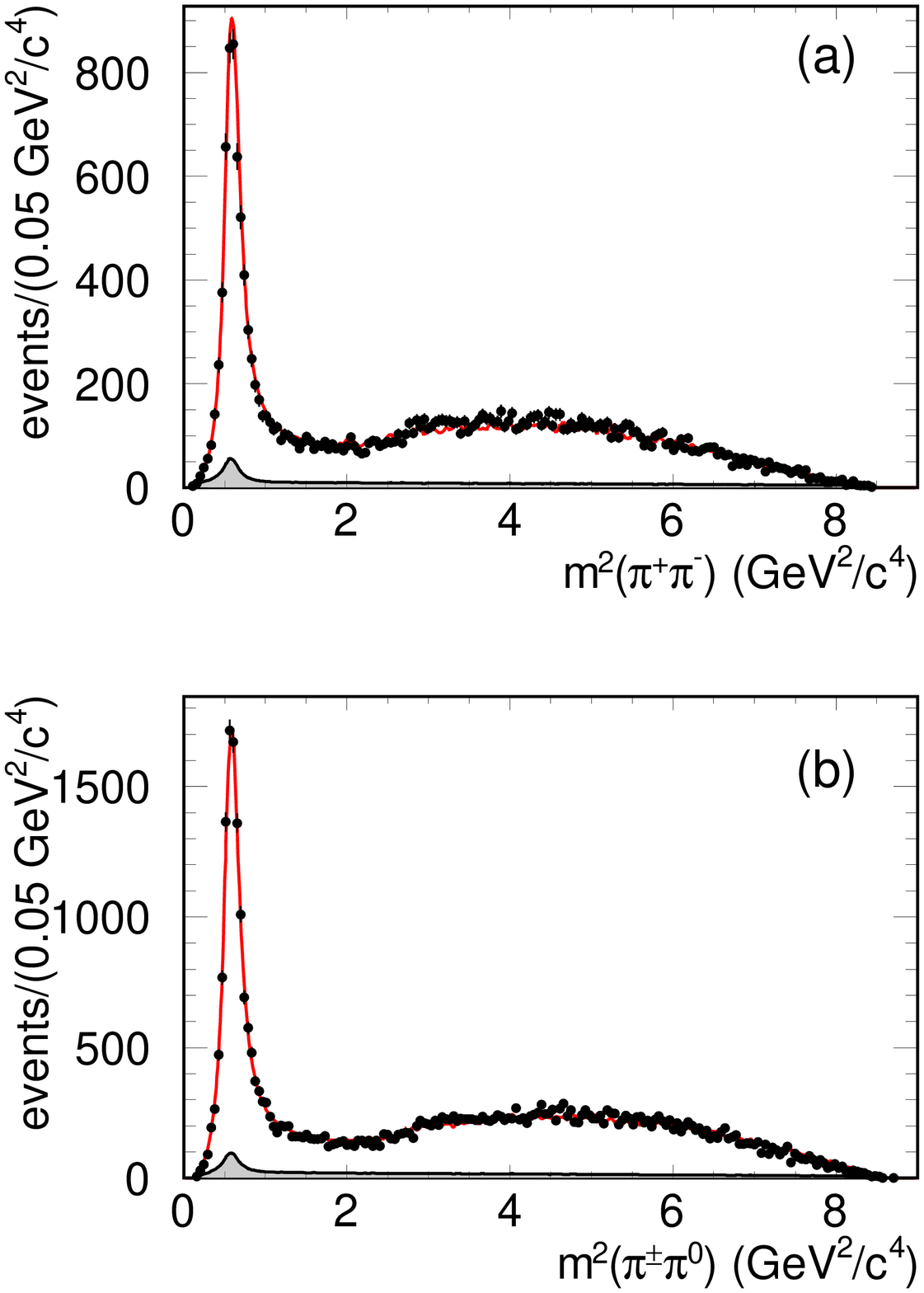}
\caption{The \psipipiz\ Dalitz plot projections. The superimposed curves result from the Dalitz-plot analysis described in the text. The shaded regions show the
background estimates obtained by interpolating the results of the Dalitz-plot analyses of the sideband regions.
}
\label{fig:fig5}
\end{center}
\end{figure}

\begin{figure}[h]
  \begin{center}
\includegraphics[width=8cm]{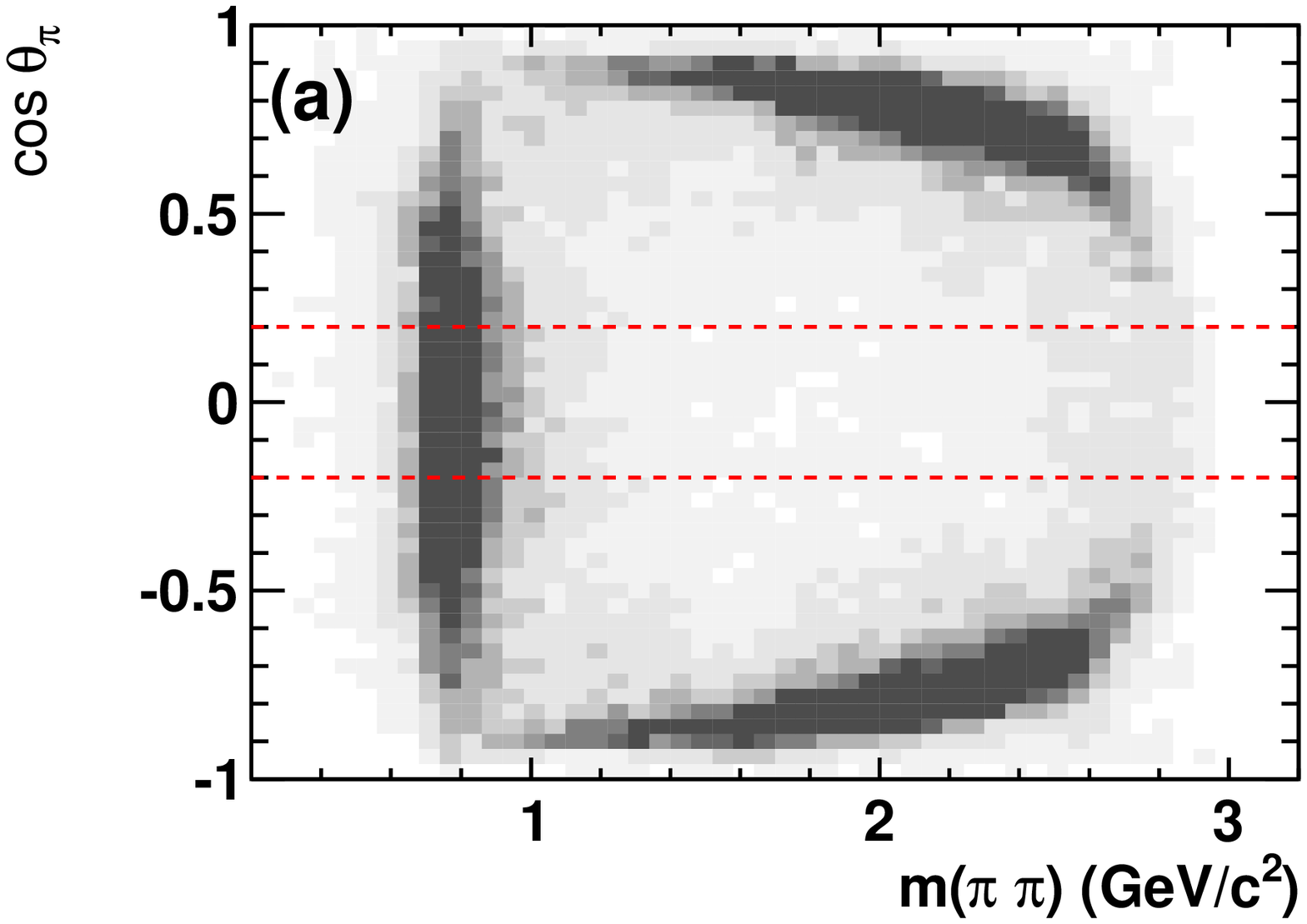}\\
\includegraphics[width=8cm]{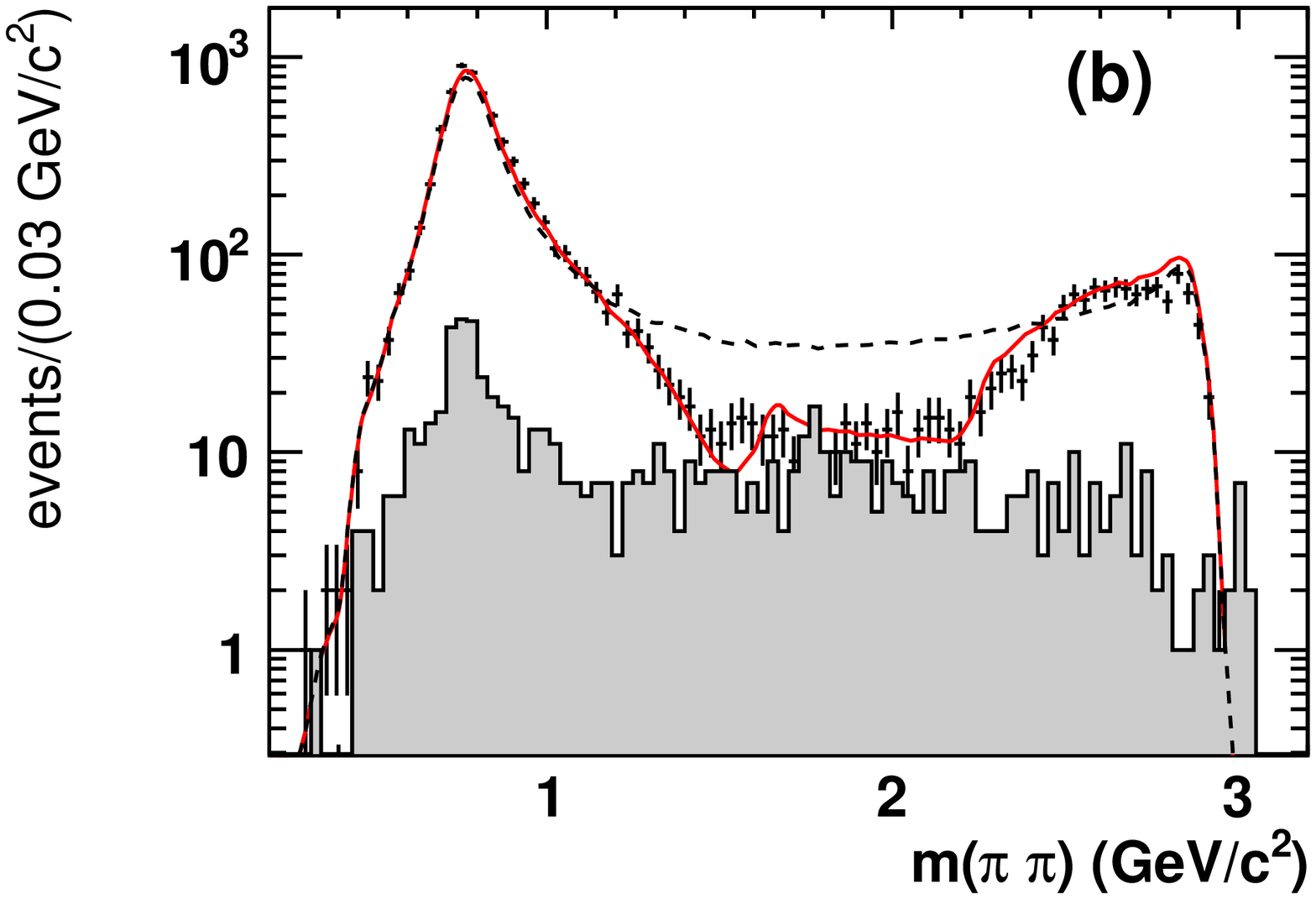}\\
\includegraphics[width=8cm]{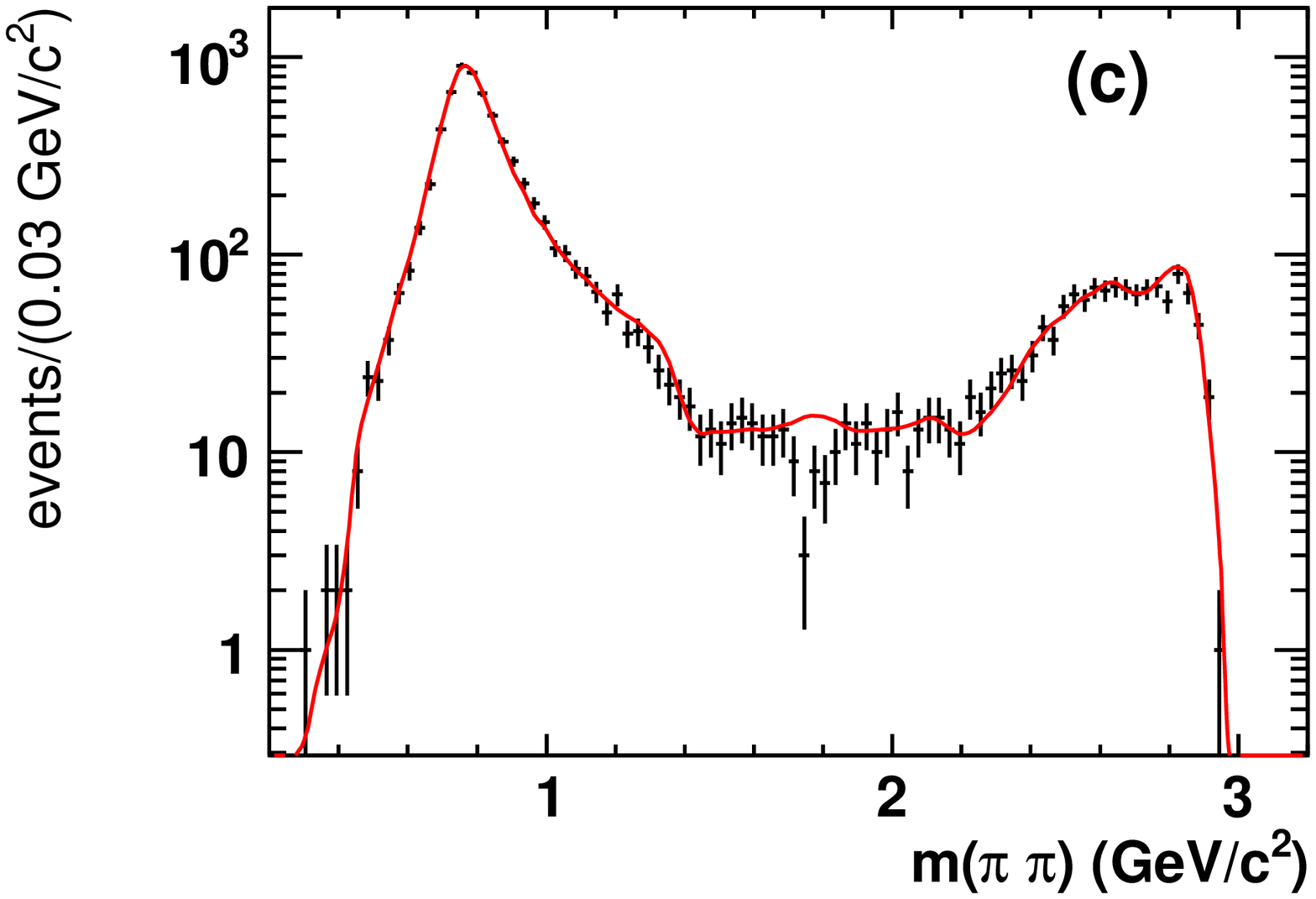}
\caption{(a) Binned scatter diagram of $\cos \theta_{\pi_3} \ vs \ m(\pi_1 \pi_2)$. (b), (c) $\pi \pi$ mass projection in the $| \cos \theta_{\pi}| < 0.2$ region for all the three $\pi \pi$ charge combinations. The horizontal lines in (a) indicate the $\cos \theta_{\pi}$ selection. The dashed line in (b) is the result from the fit with only the $\rho(770)\pi$ amplitude. The fit in (b) uses the isobar model and 
the shaded histogram shows the background distribution estimated from the \jpsi sidebands.
The fit in (c) uses the Veneziano model.}
\label{fig:fig8}
\end{center}
\end{figure}

\section{Dalitz-plot analysis}

We perform Dalitz-plot analyses of the \psipipiz, \psikkpiz, and \psikskpi candidates in the \jpsi mass region using  unbinned maximum likelihood fits.
The likelihood function is written as
\begin{eqnarray}
\mathcal{L} = \nonumber\\
 \prod_{n=1}^N&\bigg[&f_{\rm sig}(m_n) \cdot \epsilon(x'_n,y'_n)\frac{\sum_{i,j} c_i c_j^* A_i(x_n,y_n) A_j^*(x_n,y_n)}{\sum_{i,j} c_i c_j^* I_{A_i A_j^*}} \nonumber\\
& &+(1-f_{\rm sig}(m_n))\frac{\sum_{i} k_iB_i(x_n,y_n)}{\sum_{i} k_iI_{B_i}}\bigg]
\end{eqnarray}
\noindent where
\begin{itemize}
\item $N$ is the number of events in the signal region;
\item for the $n$-th event, $m_n$ is the \pipiz, \kkpiz, or \kskpi invariant mass;
\item for the $n$-th event,\\
\noindent$x_n=m^2(\pip \piz)$, $y_n=m^2(\pim \piz)$ for $\pipiz$;\\
\noindent$x_n=m^2(K^+ \piz)$, $y_n=m^2(K^- \piz)$ for $\kkpiz$; \\
\noindent$x_n=m^2(\Kpm \pimp)$, $y_n=m^2(\KS \pimp)$ for $\kskpi$;
\item $f_{\rm sig}$ is the mass-dependent fraction of signal obtained from the fits to the \pipiz, \kkpiz, and \kskpi mass spectra;
\item for the $n$-th event, $\epsilon(x'_n,y'_n)$ is the efficiency parameterized as function $x'_n=m_{12}$ and $y'_n=\cos \theta_h$ (see Sec. IV);
\item for the $n$-th event, the $A_i(x_n,y_n)$ represent the complex signal-amplitude contributions described below;
\item $c_i$ is the complex amplitude of the $i-$th signal component; the $c_i$ parameters are allowed to vary during the fit process;
\item for the $n$-th event, the $B_i(x_n,y_n)$ describe the background probability-density functions assuming that interference between signal and background amplitudes can be ignored;
\item $k_i$ is the magnitude of the $i-$th background component; the $k_i$ parameters are obtained by fitting the sideband regions;
\item $I_{A_i A_j^*}=\int A_i (x,y)A_j^*(x,y) \epsilon(m_{12},\cos \theta)\ {\rm d}x{\rm d}y$ and 
$I_{B_i}~=~\int B_i(x,y) {\rm d}x{\rm d}y$ are normalization
 integrals; numerical integration is performed on phase-space-generated events with \jpsi signal and background generated according to the experimental distributions.
\end{itemize}

Parity conservation in $\jpsi \ \to \ \pip \pim \piz$ restricts the possible spin-parity of any intermediate two-body resonance to be $J^{PC}=1^{--}, 3^{--}, ...$. 
Amplitudes are parameterized using Zemach's tensors~\cite{zemach,dionisi}. Except as noted, all fixed resonance parameters are taken
from the Particle Data Group averages~\cite{PDG}.

For reaction (\ref{eq:pi3}), we label the decay particles as
\begin{equation}
J/\psi \ \to \ \pi^+_1 \pi^-_2 \pi^0_3.
\end{equation}
Indicating with $p_i$ the momenta of the particles in the \jpsi center of mass rest frame, for a resonance $R_{jk}$ decaying as $R_{jk} \ \to \ j + k$ we also define the three-vectors $t_i$ as the vector part of
\begin{equation}
t^\mu_i = p^\mu_j - p^\mu_k - (p^\mu_j + p^\mu_k)\frac{m_j^2-m_k^2}{m_{jk}^2}.
\end{equation}
with $i,j,k$ cyclic.
We make use of the $p_i$ vectors to describe the angular momentum $L$ between $R_{jk}$ and particle $i$, and the $t_i$ vectors to describe the spin of the 
$R_{jk}$ resonance.  Since the \jpsi resonance has spin-1 and needs to be described by a vector,  the only way to obtain this result is to perform a cross-product between the $p_i$ and $t_i$ three-vectors.
Indicating with $\rho$ a generic spin-1 resonance, Table~\ref{tab:amp} reports the list of amplitudes used to the describe the \jpsi decays.
Due to Bose symmetry, the amplitudes are symmetrized with respect to the $\rho$ charge. The table also reports the expression for the nonresonant contribution 
(NR) which should also have the \jpsi quantum numbers.

For reaction (\ref{eq:kkpi0}), we label the decay particles as
\begin{equation}
J/\psi \ \to \ K^+_1 K^-_2 \pi^0_3.
\end{equation}
In this case two separate contributions are listed in Table~\ref{tab:amp}, one in which the intermediate resonance is a $K^{*\pm} \ \to \ K^{\pm} \piz$ and the other where
the intermediate resonance is a $\rho^0 \ \to \ \Kp \Km$.
The table also lists the amplitude for the $K^*_2(1430)^{\pm} K^{\mp}$
contribution. This decay mode can only occur in D-wave. To obtain this amplitude, we construct rank-2 tensors ${\bf T_i}= t_i^jt_i^k - |t_i|^2\delta^{jk}/3$ to describe the spin-2 of the $K^*_2(1430)^{\pm}$ resonance and  ${\bf P_i} = p_i^jp_i^k - |p_i|^2\delta^{jk}/3$ to describe the angular momentum
between the $K^*_2(1430)^{\pm}$ and the $K^{\mp}$. The two rank-2 tensors are then contracted into vectors ${\bf k_i}$ to obtain the spin of the $J/\psi$ resonance.
We obtain the components of ${\bf k_i}$ as $k_i^l = \sum_{\lambda=1}^{\lambda=3}{T_i^{m,\lambda}P_i^{\lambda,n}-T_i^{n,\lambda}P_i^{\lambda,m}}$ with $l,m,n$ cyclic~\cite{colangelo}. 

The amplitudes for reaction (\ref{eq:k0skpi}) are similar to those from reaction (\ref{eq:kkpi0}). In this case we label
the decay particles as
\begin{equation}
J/\psi \to K^{\pm}_1 K^0_{S2} \pi^{\mp}_3
\end{equation}
\begin{table*}[htb]
\caption{Amplitudes considered in $\jpsi \ \to \ \pipiz$, $\jpsi \ \to \ \kkpiz$ and $\jpsi \ \to \ \kskpi$ Dalitz-plot analysis. BW indicates the Breit-Wigner function.}
\label{tab:amp}
\begin{center}
\begin{tabular}{c|l|c}
\hline
\jpsi decay mode & Decay  & Amplitude \cr
\hline
  \noalign{\vskip2pt}
$\pip \pim \piz$ & $\rho \pi$  & ${\rm BW}_{\rho}(m_{13})({\bf t_2 \times p_2}) + {\rm BW}_{\rho}(m_{23})({\bf t_1 \times p_1}) + {\rm BW}_{\rho}(m_{12})({\bf t_3 \times p_3})$ \cr
 & NR & $({\bf t_1 \times p_1}) +  ({\bf t_2 \times p_2}) + ({\bf t_3 \times p_3})$ \cr
\hline
  \noalign{\vskip2pt}
$K \bar K \pi$ & $K^* \bar K$ & ${\rm BW}_{K^*}(m_{13})({\bf t_2 \times p_2}) + {\rm BW}_{K^*}(m_{23})({\bf t_1 \times p_1})$ \cr
 & $K^*_2(1430) \bar K$ & ${\rm BW}_{K^*_2}(m_{13})({\bf k_2}) + {\rm BW}_{K^*_2}(m_{23})({\bf k_1})$ \cr
 & $\rho \pi$ & ${\rm BW}_{\rho}(m_{12})({\bf t_3 \times p_3})$ \cr
\hline
\end{tabular}
\end{center}
\end{table*}

The efficiency-corrected fractional contribution $f_i$ due to resonant or nonresonant contribution $i$ is defined as follows:
\begin{equation}
f_i = \frac {|c_i|^2 \int |A_i(x_n,y_n)|^2 {\rm d}x {\rm d}y}
{\int |\sum_j c_j A_j(x,y)|^2 {\rm d}x {\rm d}y}.
    \label{eq:frac}
\end{equation}
The $f_i$ do not necessarily sum to 100\% because of interference effects. The uncertainty for each $f_i$ is evaluated by propagating the full covariance matrix obtained from the fit.

Similarly, the efficiency-corrected interference fractional contribution $f_{ij}$, for $i<j$ are defined as:
\begin{equation}
f_{ij} = \frac { \int 2 \Real[c_ic_j^* A_i(x_n,y_n) A_j(x_n,y_n)^*] {\rm d}x {\rm d}y}
{\int |\sum_j c_j A_j(x,y)|^2 {\rm d}x {\rm d}y}.
    \label{eq:frac_int}
\end{equation}

In all the Dalitz analyses described below we validate the fitting algorithms using MC simulations
with known input amplitudes and phases. We also start the fitting procedure both on MC and data from random
values. In all cases the fits converge towards one single solution.

\subsection{Dalitz-plot analysis of {\boldmath$\protect \psipipiz$}.}
\subsubsection{Isobar model.}

We perform a Dalitz-plot analysis of \psipipiz in the \jpsi signal region given in Table~\ref{tab:tab0}. This region contains 20417 events with (91.3 $\pm$ 0.2)\% purity, defined as 
$S/(S+B)$ where $S$ and $B$ indicate the number of signal and background events, respectively, as determined from the fit to the \pipiz mass spectrum shown in Fig.~\ref{fig:fig2}(a).
Sideband regions are defined as the ranges 2.919-2.980~\gevcc \ and 3.198-3.258~\gevcc, respectively.
Figure~\ref{fig:fig4} shows the Dalitz plot for the \jpsi signal region and Fig.~\ref{fig:fig5} shows the Dalitz plot projections. We observe that the decay is dominated by $\rho(770) \pi$ amplitudes which appear as non-uniform bands along the Dalitz plot boundaries.

We first perform separate fits to the \jpsi sidebands with an incoherent sum of amplitudes using the method of the channel likelihood~\cite{chafit}.
We find significant contributions from $\rho(770)$ resonances with uniform distributions of events along their bands, as well as from an incoherent uniform background.  
The resulting amplitude fractions are interpolated 
into the \jpsi signal region and normalized to the fitted purity. Figure~\ref{fig:fig5} shows the projections of the estimated background contributions as shaded 

For the description of the \jpsi Dalitz plot, amplitudes are added one at time to ascertain the associated increase of the likelihood value and decrease of the \mbox{2-D} $\chi^2$ computed on the $(m(\pip \pim),\cos\theta_h)$ plane.
We test the quality of the fit by examining a large sample of MC events at the generator level weighted 
by the likelihood fitting function and by the efficiency. These events are used to
compare the fit result to the Dalitz plot and its projections with proper normalization. The latter comparison is shown in Fig.~\ref{fig:fig5}, and good agreement is obtained for
all projections. We make use of these weighted events to compute a \mbox{2-D} $\chi^2$ over the Dalitz plot. For this purpose, we divide the Dalitz plot into a number of cells such that the expected population in each cell is at least five events. We compute 
$\chi^2 = \sum_{i=1}^{N_{\rm cells}} (N^i_{\rm obs}-N^i_{\rm exp})^2/N^i_{\rm exp}$, where $N^i_{\rm obs}$ and $N^i_{\rm exp}$ are event yields from data and simulation, respectively.

We leave free in the fit the $\rho(770)$ parameters and obtain results which are consistent with PDG
averages~\cite{PDG}.
We also leave free the $\rho(1450)$ and $\rho(1700)$ parameters in the fit and obtain a significant improvement of the Likelihood with the following resonances parameters
\begin{eqnarray}
  m(\rho(1450)) =&1429 \pm 41 \ \mevcc,
  \nonumber\\
  \Gamma(\rho(1450)) =&576 \pm 29 \ \mev, \al\al
  \nonumber\\
  m(\rho(1700)) =&1644 \pm 36 \ \mevcc,
  \nonumber\\  
  \Gamma(\rho(1700)) =&109 \pm 19 \ \mev. \al\al  
\end{eqnarray}
\begin{table*}[htb]
\caption{Results from the Dalitz-plot analysis of the \psipipiz channel. When two uncertainties are given, the first is statistical and the second
systematic. The error on the amplitude is only statistical.}
\label{tab:jpsi_pi3}
\begin{center}
\vskip -0.2cm
\begin{tabular}{lccr|c}
\hline
 \noalign{\vskip2pt}
Final state & Amplitude & Isobar fraction (\%) & Phase (radians) & Veneziano fraction (\%) \cr
\hline
 \noalign{\vskip2pt}
$\rho(770) \pi$    &     $\al1.$       & $114.2\pm1.1\al\pm2.6$  &              0. \al\al\al\al\al\al    & $133.1\pm3.3\al\al$ \cr
$\rho(1450) \pi$   & $0.513\pm0.039$   & $\al10.9\pm 1.7\al\pm2.7$   & $-2.63\pm 0.04\pm0.06$ & $0.80\pm0.27$ \cr
$\rho(1700) \pi$   & $0.067\pm0.007$   & $\al\al0.8\pm0.2\al\pm0.5$       & $-0.46\pm 0.17\pm0.21$ & $2.20\pm0.60$ \cr
$\rho(2150) \pi$   & $0.042\pm0.008$   & $\al\al0.04\pm0.01\pm0.20$    & $ \al\all1.70\pm 0.21\pm0.12$ & $6.00\pm2.50$ \cr
$\omega(783) \piz$ & $0.013\pm0.002$   & $\al\al0.08\pm0.03\pm0.02$    & $\al\all2.78\pm0.20\pm0.31$   &               \cr 
$\rho_3(1690) \pi$ &                   &                         &                        & $0.40 \pm 0.08$ \cr
 \noalign{\vskip1pt}
\hline
 \noalign{\vskip1pt}
Sum                &                   & $127.8\pm2.0\pm4.3$     &                        & $142.5\pm2.8 \al\al$ \cr
$\chi^2/\nu$       &                   &    $687/519=1.32$         &                         & $596/508=1.17\al\al$ \cr
\hline
\end{tabular}
\end{center}
\end{table*}

We also test the presence of the isospin violating decay $\omega \to \pip \pim$.
We notice that the $\omega(782) \piz$ contribution has a rather small fraction ($0.08 \pm 0.03$) but its fitted amplitude
is ($0.013 \pm  0.002$). To obtain the statistical significance for this contribution,
we remove the $\omega(782) \piz$ amplitude.
We obtain $\Delta(-2\log L)=27.7$ and $\Delta \chi^2=17$ for the difference of two parameters which corresponds to a significance of $4.9\sigma$.
We also include the spin-3 $\rho_3(1690) \pi$ contribution but it is found consistent with zero.

Table~\ref{tab:jpsi_pi3} summarizes the fit results for the amplitude fractions and phases. We note that the $\rho(770) \pi$ amplitude provides the largest contribution. We also observe an important contribution from the $\rho(1450) \pi$ amplitude, while the contribution from 
higher $\rho'$ resonances are small. We also notice that the $\rho(1700) \pi$ amplitude is significant even if the resulting fraction
is very small, which can be attributed to the presence of important interference effects.

To illustrate the contributions from higher $\rho$ states, we plot in Fig.~\ref{fig:fig8}(a), a binned scatter diagram of the 
helicity angle $\theta_{\pi_3}$ vs. $\pi_1 \pi_2$ mass for the three possible combinations. The curved bands on the top and bottom are reflections from the
other combinations. Selecting events $| \cos \theta_{\pi} | < 0.2$, almost completely removes these reflections and gives a more clear 
representation of the $\pi \pi$ mass spectrum, shown in Fig.~\ref{fig:fig8}(b) with a logarithmic scale for the sum of the three $\pi \pi$ mass combinations.
We also compare the fit projections with the results from a fit where only the $\rho(770) \pi$ contribution is included. The distribution shows
clearly the presence of higher excited $\rho$ resonances contributing to the \psipipiz decay.

The NR contribution has been included but does not improve the fit quality. The sum of the fractions
is significantly different from 100\%.
Denoting by $n\ (=8)$ the number of free parameters in the fit, we obtain $\chi^2/\nu=687/519$ ($\nu=N_{\rm cells}-n$). 

\begin{figure*}[h]
\begin{center}
\includegraphics[width=16cm]{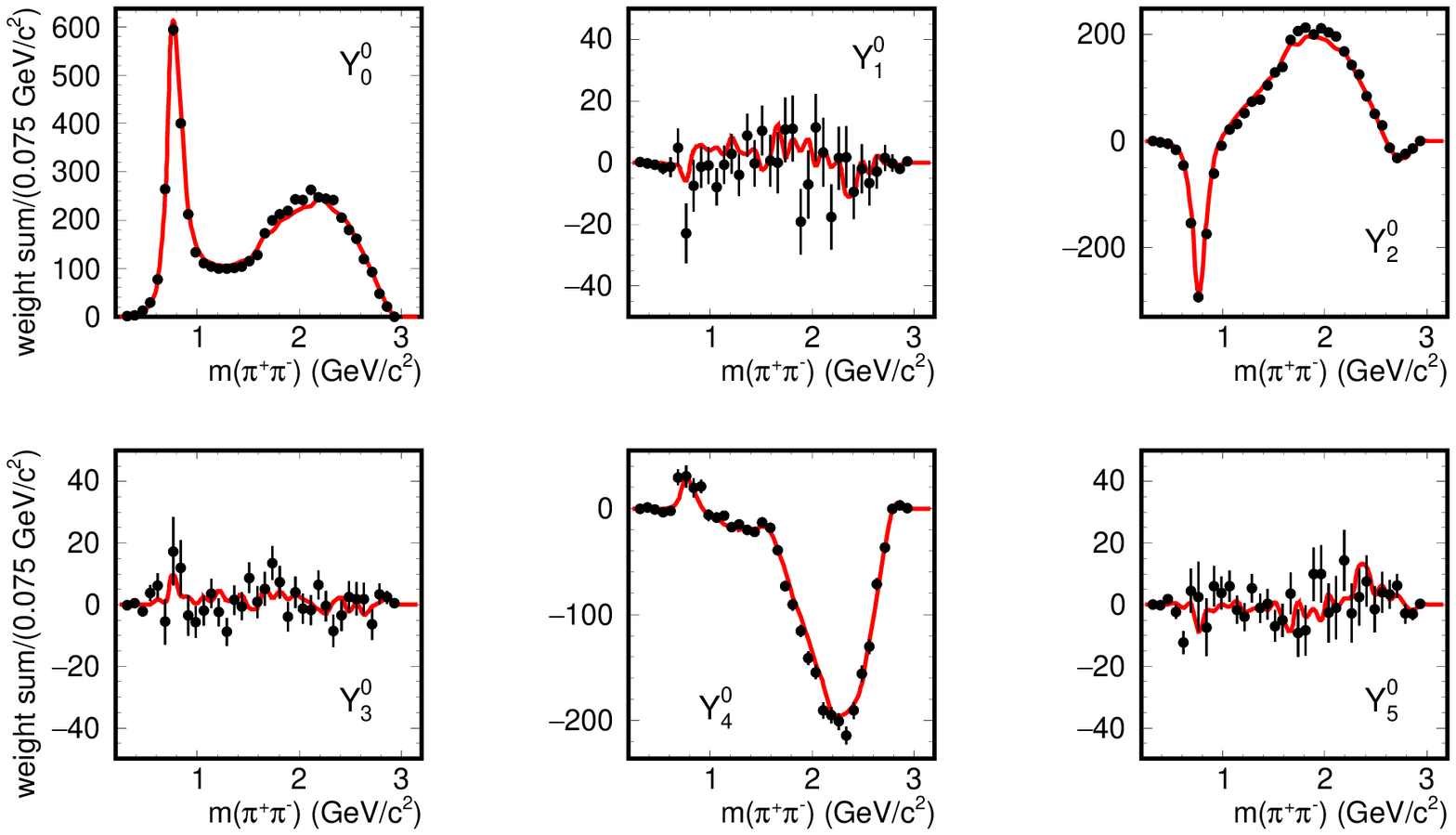}
\caption{Legendre polynomial moments for \psipipiz as a function of $\pip \pim$ mass. The superimposed curves result from the Dalitz-plot analysis described in the text.}
\label{fig:fig6}
\end{center}
\end{figure*}

\begin{figure*}[h]
\begin{center}
\includegraphics[width=16cm]{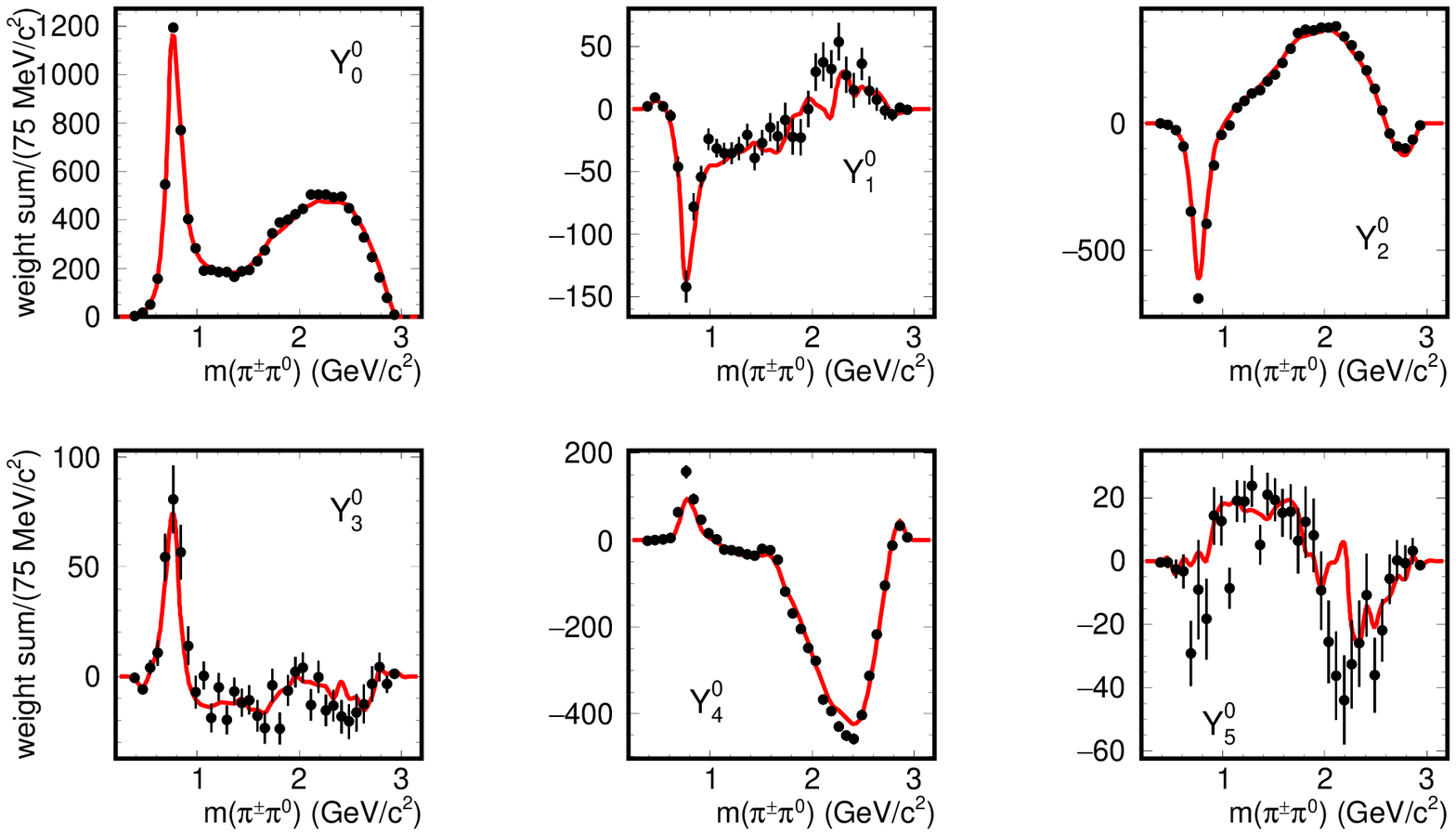}
\caption{Legendre polynomial moments for \psipipiz as a function of $\pi^{\pm} \piz$ mass. The superimposed curves result from the Dalitz-plot analysis described in the text. The corresponding $\pip \piz$ and $\pim \piz$ distributions are combined.}
\label{fig:fig7}
\end{center}
\end{figure*}

We compute the uncorrected Legendre polynomial moments $\langle Y^0_L \rangle$ in each  $\pip \pim$ and $\pi^{\pm} \piz$  mass interval by weighting each event by the relevant $Y^0_L(\cos \theta_h)$ function. These distributions are shown in Figs.~\ref{fig:fig6} and~\ref{fig:fig7}. We also compute the expected Legendre polynomial moments from the weighted MC events and compare with the experimental distributions. We observe a reasonable agreement for all the distributions, which indicates that the fit is able to reproduce most of the local structures apparent in the Dalitz plot.
We also notice a few discrepancies in the high $\pi \pi$ mass region indicating the possible presence of additional unknown excited $\rho \pi$ contributions
not included in the present analysis.

Systematic uncertainty estimates for the fractions and relative phases are computed in different ways.
\begin{itemize}
\item{i)} The purity function is scaled up and down by its statistical uncertainty.
\item{ii)} The parameters of each resonance contributing to the decay are modified within one standard deviation  
of their uncertainties in the PDG averages. 
\item{iii)} The Blatt-Weisskopf~\cite{blatt} factors entering in the relativistic Breit-Wigner function have been fixed to 1.5 $(\gevc)^{-1}$ and varied
  between 1 and 4 $(\gevc)^{-1}$.
\item{iv)} We make use of the efficiency distribution without the smoothing described in Sec.~\ref{eff}.
\item{v)} To estimate possible bias, we generate and fit MC simulated events according to the Dalitz-plot fitted results.
\end{itemize} 
The different contributions are added in quadrature in Table~\ref{tab:jpsi_pi3}.


\begin{figure}[h]
\begin{center}
\includegraphics[width=9cm]{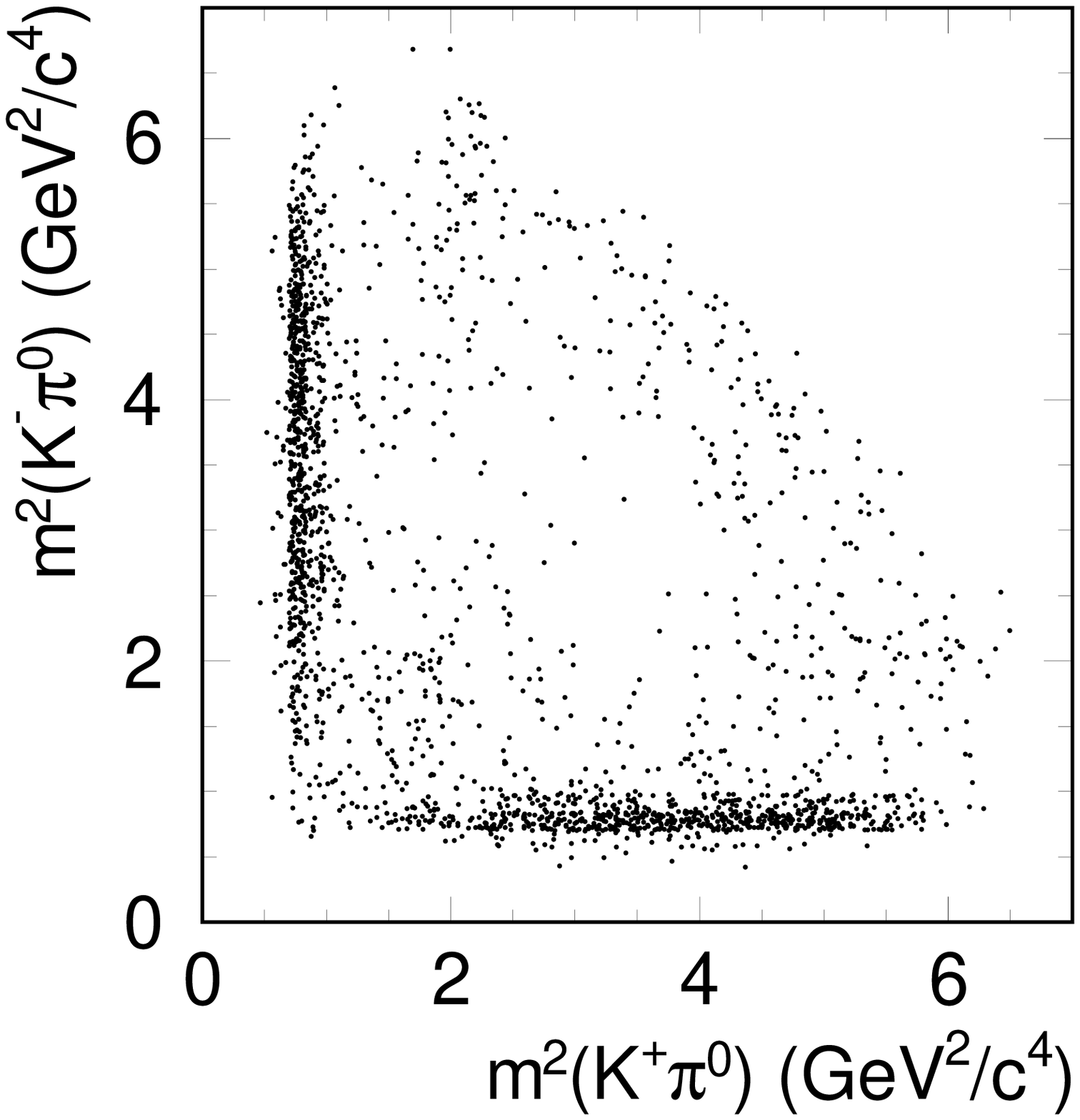}
\caption{Dalitz plot for the \psikkpiz\ events in the signal region.}
\label{fig:fig9}
\end{center}
\end{figure}

\begin{figure}[h]
\begin{center}
\includegraphics[width=9cm]{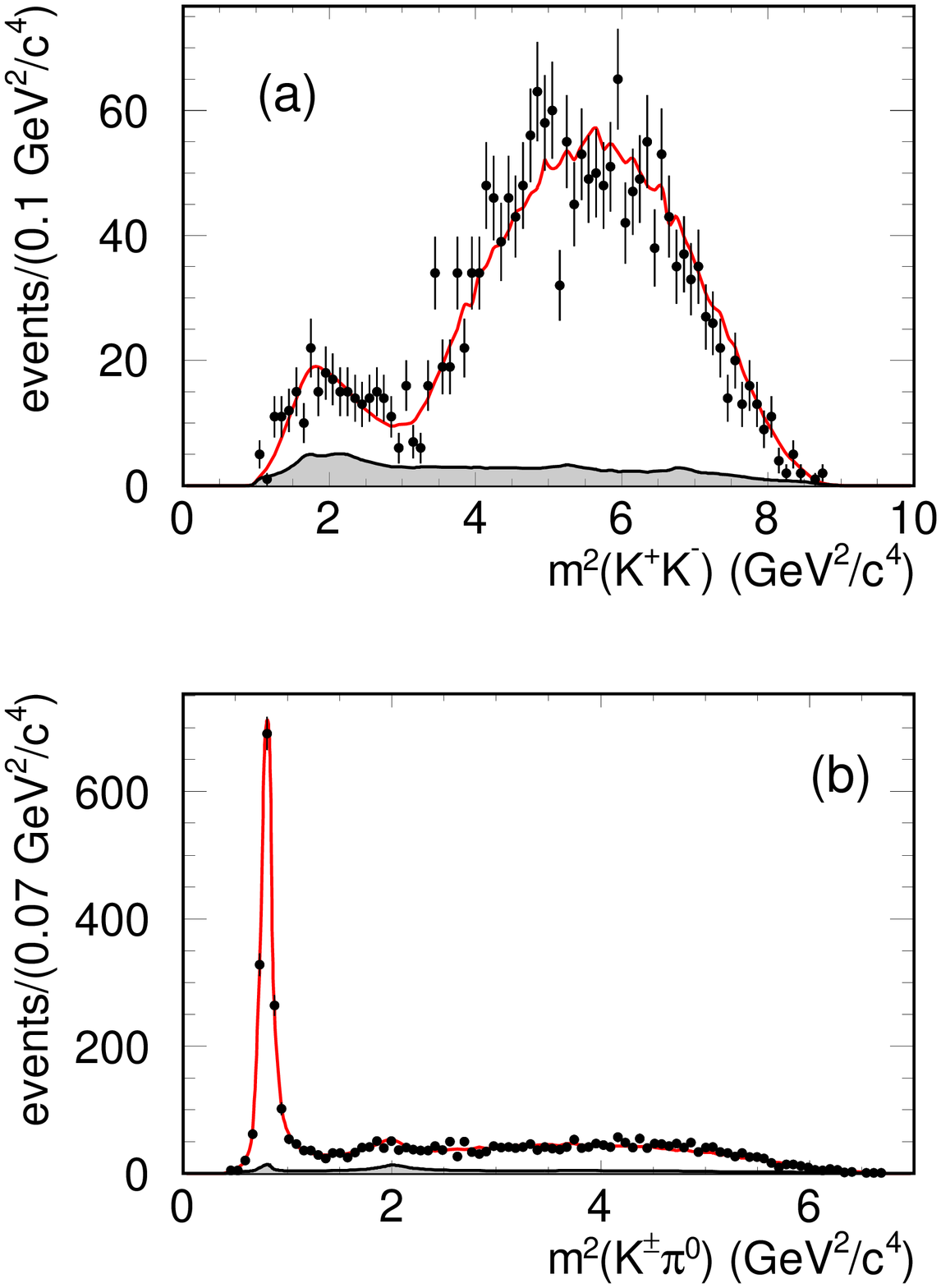}
\caption{The \psikkpiz\ Dalitz plot projections. The superimposed curves result from the Dalitz-plot analysis described in the text. The shaded regions show the
background estimates obtained by interpolating the results of the Dalitz-plot analyses of the sideband regions.
}
\label{fig:fig10}
\end{center}
\end{figure}

\begin{figure*}[h]
\begin{center}
\includegraphics[width=16cm]{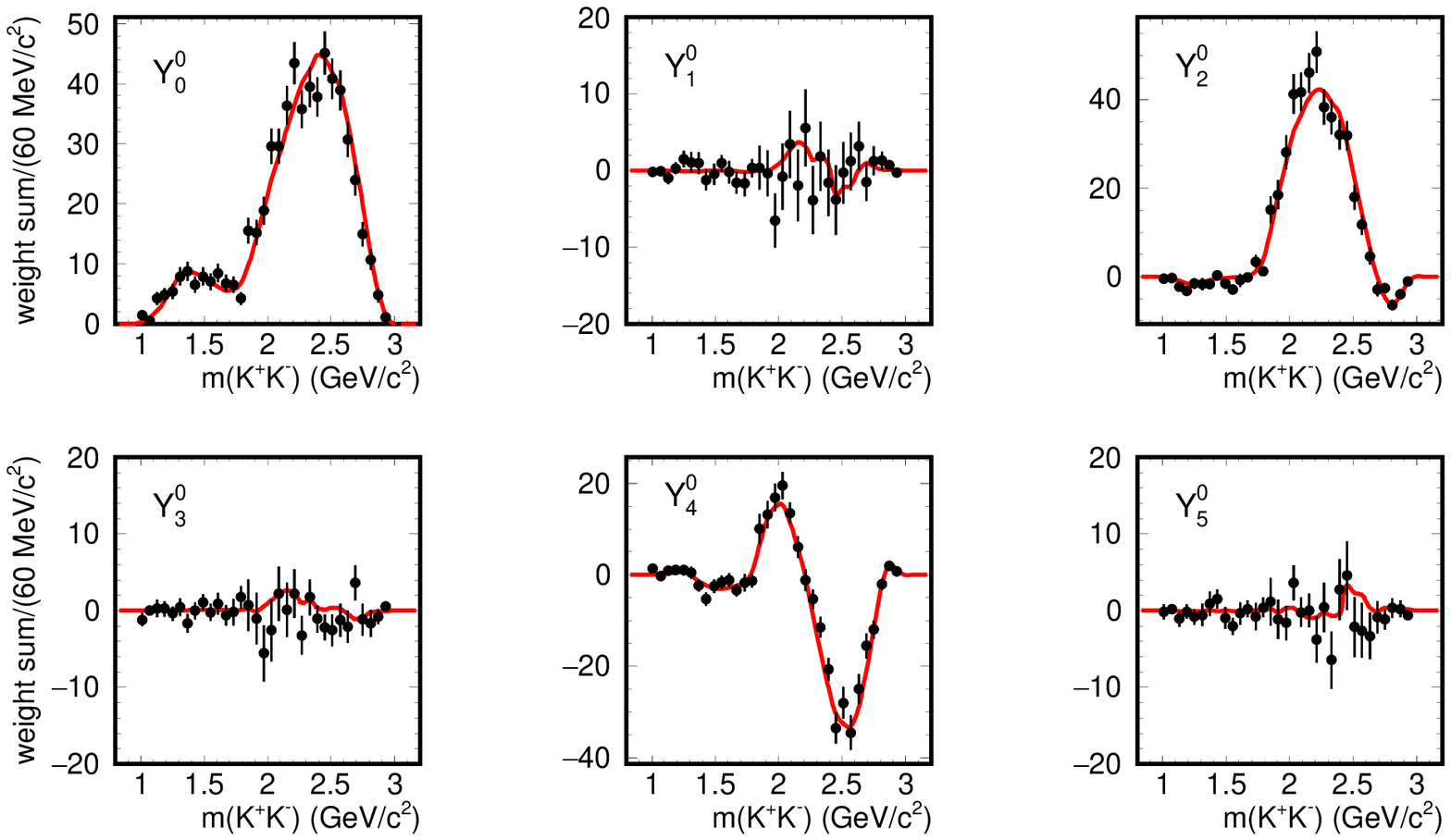}
\caption{Legendre polynomial moments for \psikkpiz as a function of $\Kp \Km$ mass. The superimposed curves result from the Dalitz-plot analysis described in the text.}
\label{fig:fig11}
\end{center}
\end{figure*}

\begin{figure*}[h]
\begin{center}
\includegraphics[width=16cm]{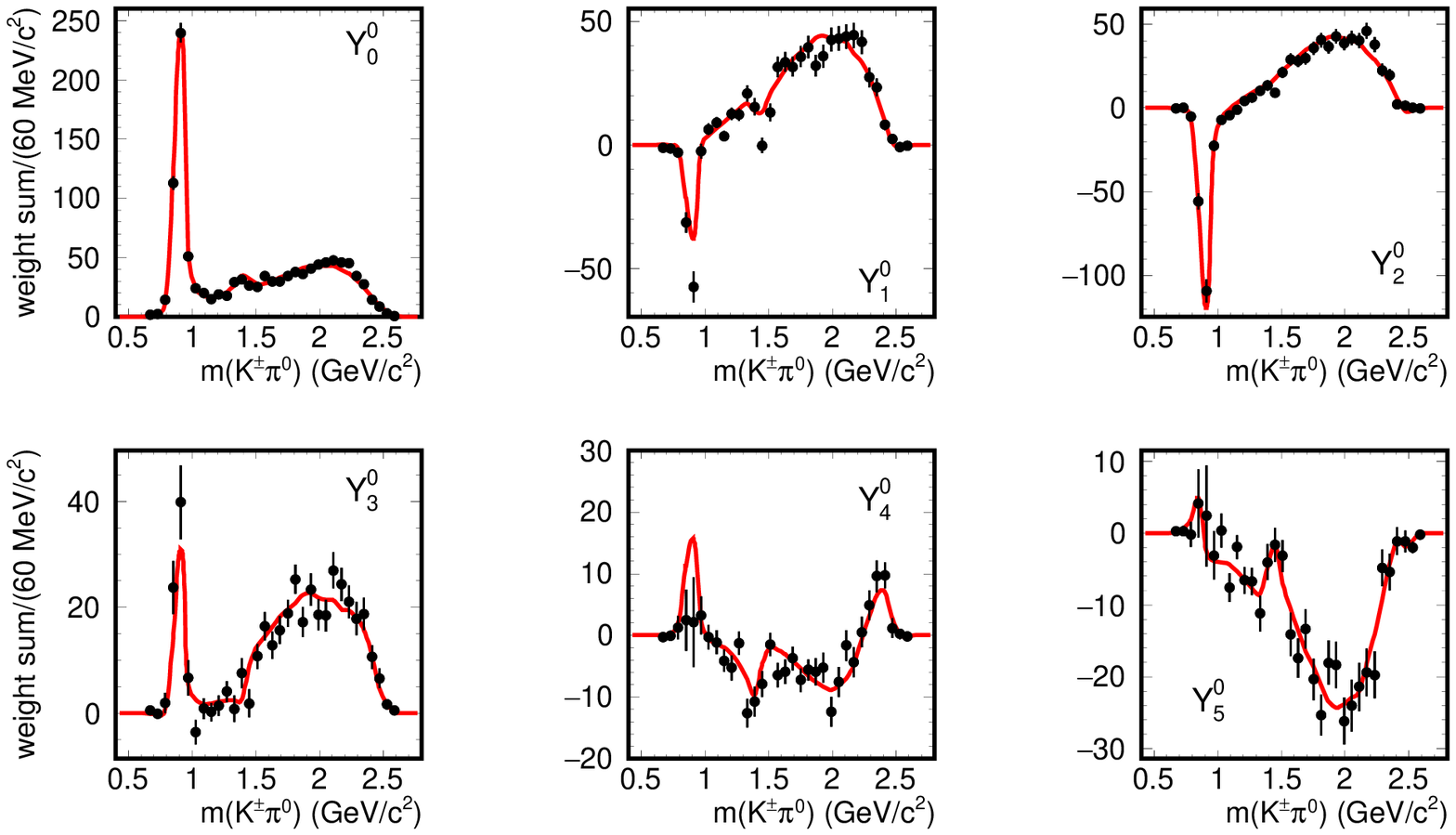}
\caption{Legendre polynomial moments for \psikkpiz as a function of $K^{\pm} \piz$ mass. The superimposed curves result from the Dalitz-plot analysis described in the text. The corresponding $\Kp \piz$ and $\Km \piz$ distributions are combined.}
\label{fig:fig12}
\end{center}
\end{figure*}

\begin{figure}[h]
\begin{center}
\includegraphics[width=9cm]{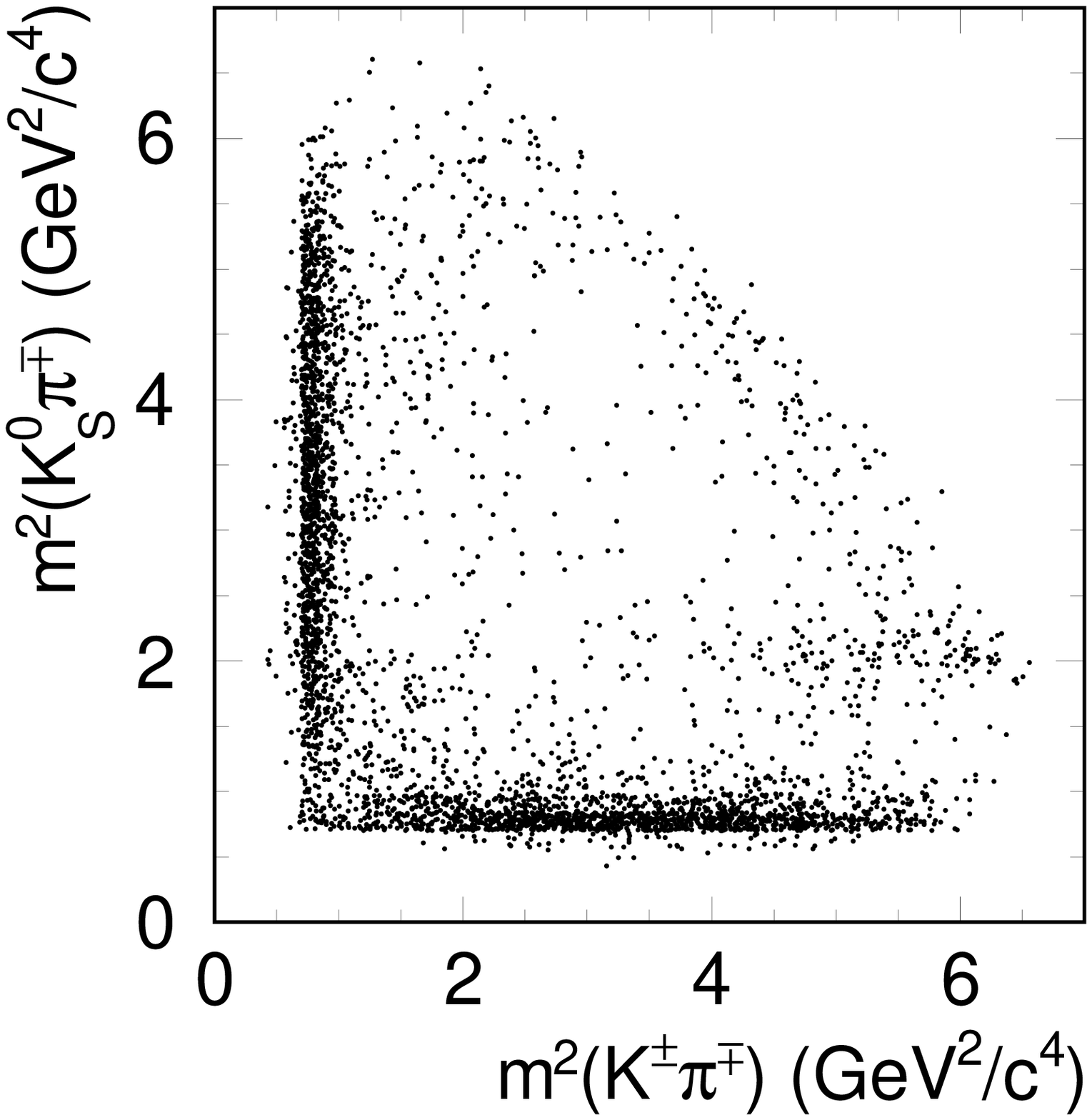}
\caption{Dalitz plot for the \psikskpi\ events in the signal region.}
\label{fig:fig13}
\end{center}
\end{figure}

\subsubsection{Veneziano model.}

The particular approach used in this analysis follows recent work described in Ref.~\cite{adam}.
The dynamical assumptions behind the Veneziano model are the resonance dominance 
of the low-energy spectrum and resonance-Regge duality. The latter means that all resonances are located
on Regge trajectories and that Regge poles are the only singularities of partial waves in the complex angular momentum
plane. Therefore, there are no ``unaccounted for'' backgrounds and the Veneziano amplitude  
  is used to fully describe the given reaction.
  A single Veneziano amplitude of the type 
   \begin{equation} 
   A_{n,m} = \frac{\Gamma(n - \alpha(s))\Gamma( n - \alpha(t))}{\Gamma(n+m - \alpha(s) - \alpha(t))}
    \label{eq:vene1}
   \end{equation} 
   has  ``predetermined'' resonance strengths.
   Here $\alpha$ is  the Regge trajectory, $s$ and $t$ are the Mandelstam variables and $n,m$ are integers.
   The position of resonances is determined by poles of the amplitude, {\it i.e.} resonances in the
   $s(t)$-channel are determined by poles of the  first (second) $\Gamma$ function in the numerator,
   respectively. Resonance couplings  are determined by residues of the amplitude at the poles.
   In the model these are therefore determined by the properties of the $\Gamma$ function and the form of
   the Regge trajectory.  Which resonances are excited depends, however, on the quantum numbers of
   external particles. Thus the amplitude in Eq.~(\ref{eq:vene1}) should be considered as a building block
   rather then a physical amplitude. The latter is obtained by forming a linear combination
   of the $A_{n,m}$'s with parameters that are reaction dependent, {\it i.e.} fitted to the data. {\it e.g.} 
\begin{equation}
A_{X\to a b c} = \sum_{n,m} c_{X \ \to \ a b c}(n,m) A_{n,m} 
    \label{eq:vene2}
\end{equation} 
In this analysis a modified set of amplitudes $A_{n,m}$, which incorporate complex trajectories
 were used. 
 Unlike the isobar model, the Veneziano model describes an infinite number of resonances. 
  The resonances are not independent, the correlation between resonance masses, $m_R$  and spins $J_R$ is described by the Regge trajectory function $\alpha(s)$  such that $\alpha(m_R^2) = J_R$. 
  Once the parameters $c$ in Eq.~(\ref{eq:vene2}) are determined by fitting data, it is possible to compute the coupling  constants of resonances to the external particles.   Weak resonances may not be 
   apparent in the data. They  however are analytically connected to other, stronger resonances and determining the latter
 helps to constrain the couplings to the former.
    For example, the $\rho_3$ meson is expected to lie on the same Regge trajectory as the $\rho$. Thus  coupling of  the $\rho$ in $J/\psi \ \to \ \rho\pi \ \to \ 3\pi$ determines coupling of the $J/\psi$ to the $\rho_3$.
    
In the Veneziano model the complexity of the model is related to $n$ which is related to the number
of Regge trajectories included in the fit. The number of free parameters also increases with $n$. 
The integer $m$, in Eq.~(\ref{eq:vene1}) is related to the number of daughter trajectories and it is restricted by 
$1 \le m \le n$. 
The lower limit on $m$ guarantees that $J/\psi$ decay amplitude has the expected high-energy behavior and the upper limit eliminates double poles in overlapping channels.
We fit the data varying  $n$ from 1 to 8 and test
the improvement in the likelihood function and the 2-D $\chi^2$. We find that no improvement is obtained with $n>7$. Taking $n=7$ the model requires
19 free parameters. Using a modified expression of Eq.~(\ref{eq:frac}) we obtain the fractions given in Table~\ref{tab:jpsi_pi3}. We observe 
a reduction of the $\rho(1450)\pi$ contribution by more than a factor of ten compared to the results from the isobar model.
However the $\rho(2150)\pi$ amplitude has a much larger contribution.
We also observe a better fit quality as compared with the isobar model.
The projection of the fit on the $\pi \pi$ mass in the $| \cos \theta_{\pi}| < 0.2$ region, is shown in Fig.~\ref{fig:fig8}(c).

We note that the isobar model gives a better description of the $\rho(1450)$ region, while the Veneziano model describes better the high mass
region. This may indicate that other resonances, apart from the low mass $\rho$ resonances, are contributing to the $J/\psi$ decay.

\subsection{Dalitz-plot analysis of {\boldmath$\protect \psikkpiz$}.}

We perform a Dalitz-plot analysis of \psikkpiz in the \jpsi signal region, defined in Table~\ref{tab:tab0}, which contains 2102 events with (88.8 $\pm$ 0.7)\% purity, as determined from the fit shown in Fig.~\ref{fig:fig2}(b).
Figure~\ref{fig:fig9} shows the Dalitz plot for the \jpsi signal region and Fig.~\ref{fig:fig10} shows the Dalitz plot projections.
We observe that the decay is dominated by the $K^*(892)^{\pm} K^{\mp}$ amplitude. We also observe a diagonal band which we tentatively attribute to the 
$\rho(1450)^0 \piz$ amplitude.

As in the previous section, we fit the
$J/\psi$ sideband regions to determine the background distribution.
Due to the limited statistics and the low background, we take enlarged sidebands, defined as the ranges 2.910-3.005~\gevcc \ and 3.176-3.271~\gevcc, respectively.
Also in this case we fit these sidebands using non-interfering amplitudes described by relativistic Breit-Wigner functions using the method of the channel likelihood~\cite{chafit}. The $K^* \bar K$ contributions are symmetrized with respect to the kaon charge. Sideband regions are dominated by the presence of $K^*(892) \bar K$ and $K^*_2(1430) \bar K$ amplitudes.

We fit the $\jpsi \ \to \ \kkpiz$ Dalitz plot using the isobar model. 
Also in this case amplitudes are added one at time to ascertain the associated increase of the likelihood value and decrease of the \mbox{2-D} $\chi^2$ 
computed on the $(m(\Kp \Km),\cos\theta_h)$ plane. The results from the best fit are summarized in Table~\ref{tab:signal_pizkk}.
We observe the following features:
\begin{itemize}
\item{} The decay is dominated by the $K^*(892)^{\pm}K^{\mp}$ and $\rho(1450)^0\piz$ amplitudes with smaller contributions from the $K^*_2(1430)^{\pm}K^{\mp}$ and $K^*_1(1410)^{\pm}K^{\mp}$ amplitudes.
\item{} We fix the $\rho(1450)$ and $\rho(1700)$ mass and width parameters to the values obtained from the $\jpsi \ \to \ \pipiz$ Dalitz plot analysis.
  This improves the description of the data, in comparison with a fit
where the masses and widths  are fixed to the PDG values~\cite{PDG}.
\item{} $K^*(1680)K$, $\rho(1700)$, $\rho(2100)$, and NR have been tried but do not give significant contributions.
\end{itemize}

We therefore assign the broad enhancement in the $\Kp \Km$ mass spectrum to the presence of the $\rho(1450)$ resonance:
the present data do not require the presence of an exotic contribution.
In evaluating the fractions we compute systematic uncertainties in a similar way as for the analysis of the $\psipipiz$ final state.
\begin{table}[htb]
\caption{Results from the Dalitz-plot analysis of the $\jpsi \ \to \ \Kp \Km \piz$ signal region. When two uncertainties are given, the first is statistical and the second systematic.}
\label{tab:signal_pizkk}
\begin{center}
\vskip -0.2cm
\begin{tabular}{lcc}
\hline
Final state & fraction (\%) & phase (radians)\cr
\hline
  \noalign{\vskip2pt}
$K^*(892)^{\pm} K^{\mp}$ & $92.4 \pm 1.5 \pm 3.4\al$ & 0. \cr
$\rho(1450)^0 \piz$ & $9.3 \pm 2.0 \pm 0.6$ & $\al3.78 \pm 0.28 \pm 0.08$\cr
$K^*(1410)^{\pm} K^{\mp}$ & $2.3 \pm 1.1 \pm 0.7$ & $\al3.29 \pm 0.26 \pm 0.39$\cr
$K^*_2(1430)^{\pm} K^{\mp}$ & $3.5 \pm 1.3 \pm 0.9$ & $-2.32 \pm 0.22 \pm 0.05\all$\cr
\hline
Total & $107.4 \pm 2.8$   & \cr
$\chi^2/\nu$  & $132/137=0.96$ & \cr
\hline
\end{tabular}
\end{center}
\end{table}

We compute the uncorrected Legendre polynomial moments $\langle Y^0_L \rangle$ in each  $\Kp \Km$ and $K^{\pm} \piz$  mass interval by weighting each event by the relevant $Y^0_L(\cos \theta)$ function. These distributions are shown in Figs.~\ref{fig:fig11} and~\ref{fig:fig12}. We also compute the expected Legendre polynomial moments from the weighted MC events and compare these with the experimental distributions. We observe good agreement for all the distributions, which indicates that the fit is able to reproduce the local structures apparent in the Dalitz plot.

\subsection{Dalitz-plot analysis of {\boldmath$\protect \psikskpi$}.}
\begin{figure}[h]
\begin{center}
\includegraphics[width=9cm]{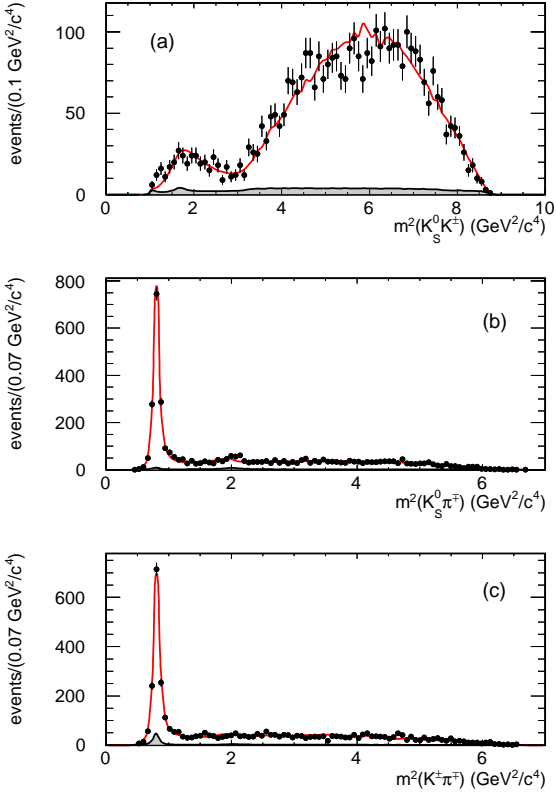}
\caption{The \psikskpi\ Dalitz plot projections. The superimposed curves result from the Dalitz-plot analysis described in the text. The shaded regions show the
background estimates obtained by interpolating the results of the Dalitz-plot analyses of the sideband regions.
}
\label{fig:fig14}
\end{center}
\end{figure}

\begin{figure*}[h]
\begin{center}
\includegraphics[width=16cm]{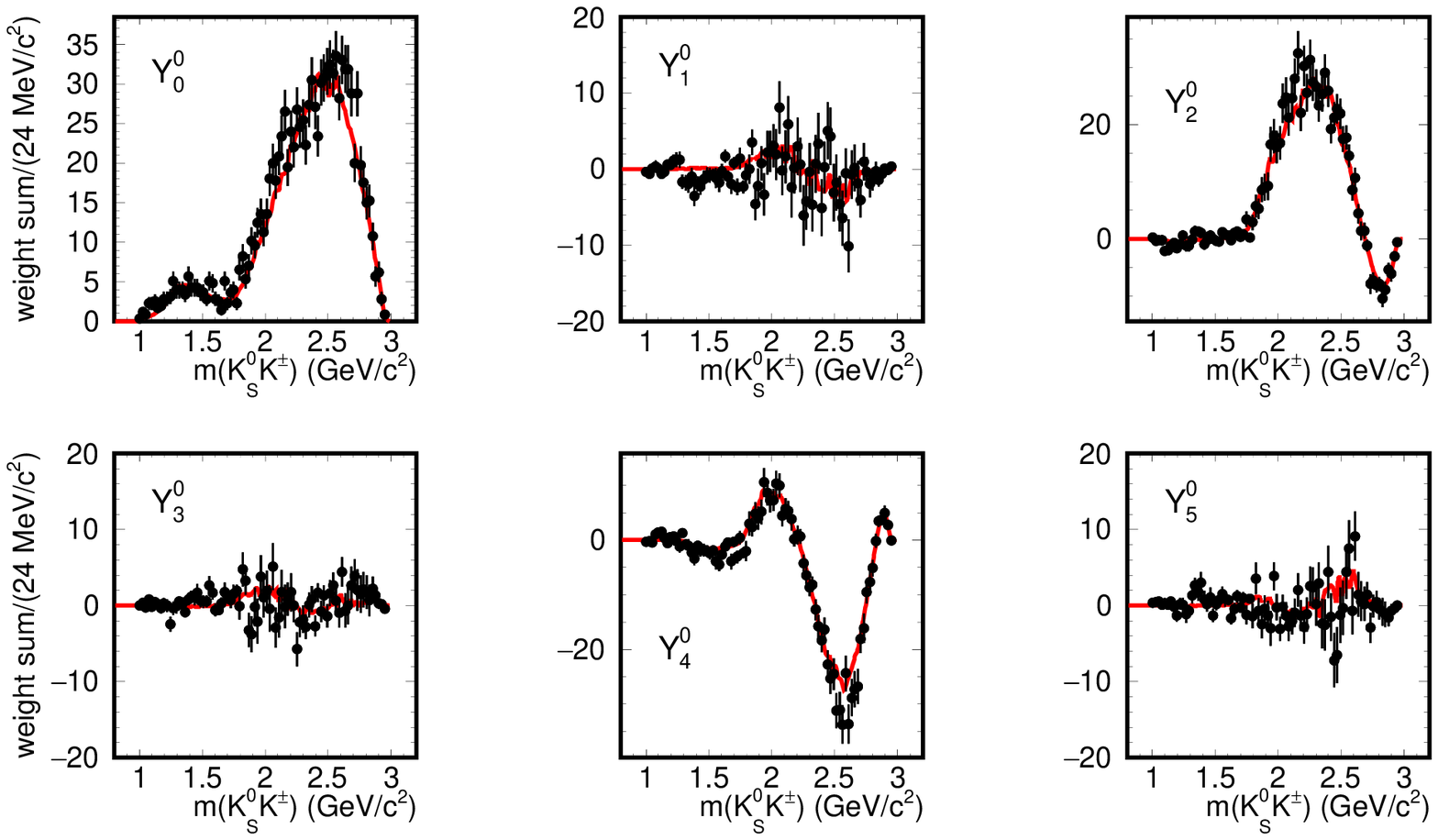}
\caption{Legendre polynomial moments for \psikskpi as a function of $\KS \Kpm$ mass. The superimposed curves result from the Dalitz-plot analysis described in the text.}
\label{fig:fig15}
\end{center}
\end{figure*}

\begin{figure*}[h]
\begin{center}
\includegraphics[width=16cm]{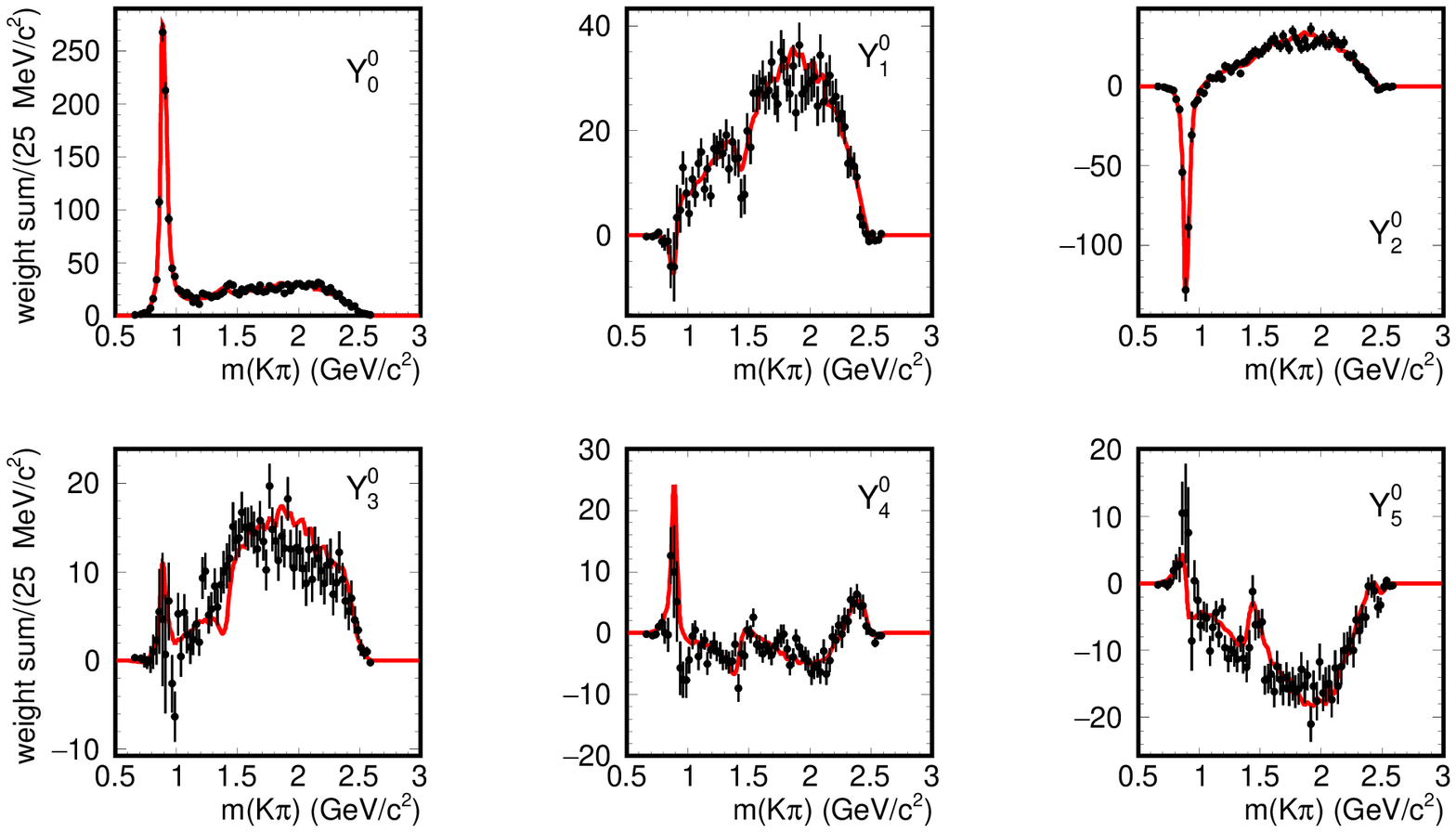}
\caption{Legendre polynomial moments for \psikskpi as a function of $K \pi$ mass. The superimposed curves result from the Dalitz-plot analysis described in the text. The corresponding $\KS \pimp$ and $\Kpm \pimp$ distributions are combined.}
\label{fig:fig16}
\end{center}
\end{figure*}

We perform a Dalitz plot analysis of \psikskpi in the \jpsi signal region defined in Table~\ref{tab:tab0}. This region contains 3907 events with (93.1 $\pm$ 0.4)\% purity, as determined from the fit shown in Fig.~\ref{fig:fig2}(c).
Figure~\ref{fig:fig13} shows the Dalitz plot for the \jpsi signal region and Fig.~\ref{fig:fig14} shows the Dalitz plot projections.

As in the previous sections, we fit the
$J/\psi$ sideband regions to determine the background distribution using the channel likelihood~\cite{chafit} method.

We fit the $\jpsi \ \to \ \kskpi$ Dalitz plot using the isobar model. 
Amplitudes have been included one by one
testing the likelihood values and the 2-D $\chi^2$ computed on the $(m(\KS \Kpm),\cos\theta_h)$ plane.
The results from the best fit are summarized in Table~\ref{tab:signal_k0skpi}.
We observe the following features:
\begin{itemize}
\item{} The decay is dominated by the $K^*(892) \bar K$, $K^*_2(1430) \bar K$  and $\rho(1450)^{\pm}\pi^{\mp}$ amplitudes with a smaller contribution from the
  $K^*_1(1410) \bar K$ amplitude.
\item{} We obtain a significant improvement of the description of the data by leaving free the $K^*(892)$ mass and width parameters and obtain
\begin{eqnarray}
m(K^*(892)^+) =& 895.6 \pm 0.8 \ \mevcc,
  \nonumber\\
\Gamma(K^*(892)^+) =& 43.6 \pm 1.3 \ \mev, \al\all
  \nonumber\\
m(K^*(892)^0) =& 898.1 \pm 1.0 \ \mevcc,
  \nonumber\\  
\Gamma(K^*(892)^0) =& 52.6 \pm 1.7 \ \mev. \al\al
\end{eqnarray}
The measured parameters for the charged $K^*(892)^+$ are in good agreement with those measured in $\tau$ lepton decays~\cite{PDG}.
\item{} We fix the $\rho(1450)$ and $\rho(1700)$ parameters to the values obtained from the $\jpsi \ \to \ \pipiz$ Dalitz plot analysis.
  This improves the description of the data in comparison with a fit
where the masses and widths  are fixed to the PDG values~\cite{PDG}.
\item{} $K^*(1680) \bar K$, $\rho(1700) \pi$, $\rho(2100) \pi$, and NR amplitudes have been tried but do not give significant contributions.
\end{itemize}

We therefore assign the broad enhancement in the $\KS \Kpm$ mass spectrum to the presence of the $\rho(1450)^{\pm}$ resonance.
In evaluating the fractions we compute systematic uncertainties in a similar way as for the analysis of the $\psipipiz$ and \psikkpiz final states.
\begin{table}[htb]
\caption{Results from the Dalitz-plot analysis of the $\jpsi \ \to \ \KS \Kpm \pimp$ signal region. When two uncertainties are given, the first is statistical and the second
systematic.}
\label{tab:signal_k0skpi}
\begin{center}
\vskip -0.2cm
\begin{tabular}{lcc}
\hline
Final state & fraction (\%) & phase (radians)\cr
\hline
  \noalign{\vskip2pt}
$K^*(892) \bar K$ & $90.5 \pm 0.9 \pm 3.8\al$ & 0. \cr
$\rho(1450)^{\pm} \pi^{\mp}$ & $6.3 \pm 0.8 \pm 0.6$ & $-3.25 \pm 0.13 \pm 0.21$\cr
$K^*_1(1410) \bar K$ & $1.5 \pm 0.5 \pm 0.9$ & $\al\all1.42 \pm 0.31 \pm 0.35$\cr
$K^*_2(1430) \bar K$ & $7.1 \pm 1.3 \pm 1.2$ & $-2.54 \pm 0.12 \pm 0.12$\cr
\hline
Total & $105.3 \pm 3.1$  & \cr
\hline
$\chi^2/\nu$ & $274/217=1.26$ & \cr
\hline
\end{tabular}
\end{center}
\end{table}
We compute the uncorrected Legendre polynomial moments $\langle Y^0_L \rangle$ in each  $\KS \Kpm$,
$K^{\pm} \pimp$, and $\KS \pimp$ mass interval by weighting each event by the relevant $Y^0_L(\cos \theta)$ function. These distributions are shown in Fig.~\ref{fig:fig15} as functions of the $\KS K^{\pm}$ mass and in Fig.~\ref{fig:fig16} as functions of the $K \pi$ mass, combining the
$\KS \pi^{\mp}$ and $K^{\pm}\pi^{\mp}$ distributions. We also compute the expected Legendre polynomial moments from the weighted MC events and compare them with the experimental distributions. We observe good agreement for all the distributions, which indicates that the fit is able to reproduce the local structures apparent in the Dalitz plot.

\section{Measurement of the $\rho(1450)^0$ relative branching fraction.}

In the Dalitz-plot analysis of \psikkpiz, the data are consistent with the observation of the decay $\rho(1450)^0 \ \to \ \Kp \Km$.
This allows a measurement of its relative branching fraction to $\rho(1450)^0 \ \to \ \pip \pim$.

We notice that the Veneziano model gives a $\rho(1450)$ contribution which is ten times smaller than the isobar model. No equivalent Veneziano analysis of the $\jpsi \ \to \ K^+ K^- \pi^0$ decay has been performed, therefore  we perform a measurement of the $\rho(1450)$ relative branching fraction using the isobar model only.

We have measured in Sec. V (Eq.~(\ref{eq:rr1})) the ratio $\calR~=~\BR(\jpsi \ \to \ K^+ K^- \pi^0)/\BR(\jpsi \ \to \ \pi^+ \pi^- \pi^0)$ and obtain $\calR=0.120 \pm 0.003 \pm 0.009$.
From the Dalitz-plot analysis of $J/\psi \ \to \ \pip \pim \piz$ and $J/\psi \ \to \ \Kp \Km \piz$ we obtain the $\rho(1450)^0$ fractions
whose systematic uncertainties are found to be independent.
From the Dalitz-plot analysis of $J/\psi \ \to \ \pip \pim \piz$ we obtain:
\begin{equation}
\begin{split}
\calB_1 = &\frac{\BR(J/\psi \ \to \ \rho(1450)^0 \piz)\BR(\rho(1450)^0 \to \pip \pim)}{\BR(\jpsi \ \to \ \pi^+ \pi^- \pi^0)}  \\
= & [(10.9 \pm 1.7({\rm stat}) \pm 2.7({\rm sys}))/3.]\% \\
= &(3.6 \pm 0.6({\rm stat}) \pm 0.9({\rm sys}))\%.
\end{split}
\end{equation}
From the Dalitz-plot analysis of $J/\psi \ \to \ \Kp \Km \piz$ we obtain:
\begin{equation}
\begin{split}
\calB_2 = &\frac{\BR(J/\psi \ \to \ \rho(1450)^0 \piz)\BR(\rho(1450)^0 \to \Kp \Km)}{\BR(\jpsi \ \to \ \Kp \Km \piz)}  \\
 = & (9.3 \pm  2.0({\rm stat}) \pm 0.6({\rm sys}))\%.
\end{split}
\end{equation}
We therefore obtain:
\begin{equation}
\begin{split}
\frac{\BR(\rho(1450)^0 \ \to \ \Kp \Km)}{\BR(\rho(1450)^0 \ \to \ \pip \pim)} \\
= &\frac{\calB_2}{\calB_1}\cdot \calR\\
= &0.307 \pm 0.084({\rm stat}) \pm 0.082({\rm sys}).
\end{split}
\end{equation}

\section{Summary}
We study the processes $\epem \ \to \ \gamma_{\rm ISR} \jpsi$ where \psipipiz, \psikkpiz, and \psikskpi using a data sample
of 519~\invfb\ recorded with the \babar\ detector operating at the SLAC PEP-II
asymmetric-energy \epem\ collider at center-of-mass energies at and near the
$\Upsilon(nS)$ ($n = 2,3,4$) resonances. 
We measure the branching fractions: 
$\calR_1 = \frac{\BR(J/\psi \ \to \ K^+ K^- \pi^0)}{\BR(J/\psi \ \to \ \pi^+ \pi^- \pi^0)} = 0.120 \pm 0.003({\rm stat}) \pm 0.009({\rm sys})$, and
$\calR_2 = \frac{\BR(\jpsi \ \to \KS K^{\pm} \pi^{\mp})}{\BR(\jpsi \ \to \ \pip \pim \piz)} = 0.265 \pm 0.005 ({\rm stat})\pm 0.021({\rm sys})$.
We perform Dalitz-plot analyses of the three \jpsi decay modes and measure fractions for resonances contributing to the decays. 
We also perform a Dalitz-plot analysis of \psipipiz using the Veneziano model. We observe structures compatible with the presence
of $\rho(1450)^0$ in both \psipipiz and \psikkpiz and measure the ratio of branching fractions:
$\calR(\rho(1450)^0) = \frac{\BR(\rho(1450)^0 \ \to \ K^+ K^-)}{\BR(\rho(1450)^0 \ \to \ \pi^+ \pi^-)} = 0.307 \pm 0.084 ({\rm stat})\pm 0.082({\rm sys})$

\section{Acknowledgements}
We are grateful for the 
extraordinary contributions of our \pep2\ colleagues in
achieving the excellent luminosity and machine conditions
that have made this work possible.
The success of this project also relies critically on the 
expertise and dedication of the computing organizations that 
support \babar.
The collaborating institutions wish to thank 
SLAC for its support and the kind hospitality extended to them. 
This work is supported by the
US Department of Energy
and National Science Foundation, the
Natural Sciences and Engineering Research Council (Canada),
the Commissariat \`a l'Energie Atomique and
Institut National de Physique Nucl\'eaire et de Physique des Particules
(France), the
Bundesministerium f\"ur Bildung und Forschung and
Deutsche Forschungsgemeinschaft
(Germany), the
Istituto Nazionale di Fisica Nucleare (Italy),
the Foundation for Fundamental Research on Matter (The Netherlands),
the Research Council of Norway, the
Ministry of Education and Science of the Russian Federation,
Ministerio de Economia y Competitividad (Spain), and the
Science and Technology Facilities Council (United Kingdom).
Individuals have received support from 
the Marie-Curie IEF program (European Union), the A. P. Sloan Foundation (USA) 
and the Binational Science Foundation (USA-Israel).
The work of A.Palano, M.R.Pennington and A.P.Szczepaniak, were supported (in part) by the U.S. Department of Energy, Office of Science, Office of Nuclear Physics under contract DE-AC05-06OR23177.
We acknowledge P. Colangelo for useful suggestions.

\section{APPENDIX}

The central values and statistical errors for the interference fit fractions are shown in Table~\ref{tab:int_pipi0}, Table~\ref{tab:int_kkpi0}, and Table~\ref{tab:int_k0skpi}, for the \psipipiz, \psikkpiz, and \psikskpi, respectively.
Table~\ref{tab:vene} reports the fitted  $c_{X \ \to \ a b c}(n,m)$ coefficients with statistical uncertainties from the Veneziano model description of \psipipiz.

\begin{table*}
\caption{\small Interference fit fractions (\%) and statistical uncertainties from the Dalitz plot analysis of \psipipiz. The amplitudes are: 
($A_0$) $\rho(770) \pi$, ($A_1$) $\rho(1450) \pi$, ($A_2$) $\rho(1700) \pi$, ($A_3$) $\rho(2150) \pi$, ($A_4$) $\omega(783) \piz$. The diagonal elements are the same as the conventional fit fractions.
}
\centering
\vspace{1ex}
\begin{tabular}{lccccc}
\hline
& $A_0$ & $A_1$ & $A_2$ & $A_3$ & $A_4$ \\
\hline
$A_0$ & $114.2\pm1.1$ & $-10.4\pm0.8$ & \al\all$0.7\pm0.1$ & \al\all$0.1\pm0.1$ & $-1.1\pm0.3$ \\
$A_1$ &  & $\al\all10.9\pm 1.7$ & $-1.7\pm0.6$ & $-0.2\pm0.1$ & \al\all$0.0\pm0.0$ \\
$A_2$ &  &  & $\al\all0.8\pm0.2$ & $-0.07\pm0.02$ & \al\all$0.0\pm0.0$ \\
$A_3$ &  &  &  & \al\all$0.04\pm0.01$ & \al\all$0.0\pm0.0$ \\
$A_4$ &  &  &  &  & \al\all$0.08\pm0.03$ \\
\hline
\end{tabular}
\label{tab:int_pipi0}
\end{table*}

\begin{table*}
\caption{\small Interference fit fractions (\%) and statistical uncertainties from the Dalitz plot analysis of \psikkpiz. The amplitudes are: 
($A_0$) $K^*(892)^{\pm} K^{\mp}$, ($A_1$) $\rho(1450)^0 \piz$, ($A_2$) $K^*(1410)^{\pm} K^{\mp}$, ($A_3$) $K^*_2(1430)^{\pm} K^{\mp}$. The diagonal elements are the same as the conventional fit fractions.
}
\centering
\vspace{1ex}
\begin{tabular}{lcccc}
\hline
& $A_0$ & $A_1$ & $A_2$ & $A_3$ \\
\hline
$A_0$ & \all$92.4\pm1.5$ & $-5.5\pm0.6$ & $-0.7\pm0.1$ & $-0.9\pm0.2$\\
$A_1$ &  & \all\all\al$9.3\pm 2.0$ &$\al\al2.2\pm0.7$ & \al\al$2.1\pm0.4$ \\
$A_2$ &  &  & \al\al$2.3\pm1.1$ & \al\al$3.3\pm0.9$ \\
$A_3$ &  &  &  & \al\al$3.5\pm1.3$ \\
\hline
\end{tabular}
\label{tab:int_kkpi0}
\end{table*}

\begin{table*}
\caption{\small Interference fit fractions (\%) and statistical uncertainties from the Dalitz plot analysis of $\jpsi \ \to \ \KS \Kpm \pimp$. The amplitudes are: 
($A_0$) $K^*(892) \bar K$, ($A_1$) $\rho(1450)^{\pm} \pi^{\mp}$, ($A_2$) $K^*_1(1410) \bar K$, ($A_3$) $K^*_2(1430) \bar K$. The diagonal elements are the same as the conventional fit fractions.
}
\centering
\vspace{1ex}
\begin{tabular}{lcccc}
\hline
& $A_0$ & $A_1$ & $A_2$ & $A_3$ \\
\hline
$A_0$ & \all$90.5\pm0.9$ & \all$-5.4\pm0.4$ & \al\al$0.1\pm0.1$ & \all$-1.3\pm0.2$\\
$A_1$ &  & \all\all\al$6.3\pm 0.8$ &$\all-0.1\pm0.5$ & \al\al$1.9\pm0.3$ \\
$A_2$ &  &  & \al\al$1.5\pm0.5$ & \al\al\all$3.3\pm1.6$ \\
$A_3$ &  &  &  & \al\al$7.1\pm1.3$ \\
\hline
\end{tabular}
\label{tab:int_k0skpi}
\end{table*}

\begin{table*}
\caption{\small Fitted $c_{X \ \to \ a b c}(n,m)$ coefficients with statistical uncertainties from the Veneziano model description of \psipipiz.
}
\centering
\vspace{1ex}
\begin{tabular}{c|c| r@{}l}
\hline
 $n$ & $m$ & \multicolumn{2}{c} {$c_{X \ \to \ a b c}(n,m)$} \\
\hline
 1 & 1 &  0.5720 $\pm$ & \, 0.0016 \cr
 2 & 1 &  0.7380 $\pm$ & \,  0.0027 \cr
 3 & 1 &  0.1165 $\pm$ & \, 0.0014 \cr
   & 2 &  4901  $\pm$ & \, 426 \cr
 4 & 1 &  354  $\pm$ & \, 53 \cr
   & 2 &  1781  $\pm$ & \, 49 \cr
 5 & 1 &  -137.4  $\pm$ & \,3.4 \cr
   & 2 &  2087  $\pm$ & \,245 \cr
   & 3 &  -248  $\pm$ & \,25 \cr
 6 & 1 &  1869  $\pm$ & \,86 \cr
   & 2 &  -354  $\pm$ & \,10 \cr
   & 3 &  9.8  $\pm$ & \,0.3 \cr
 7 & 1 &  1084  $\pm$ & \,132 \cr
   & 2 &  63.5  $\pm$ & \,13.7 \cr
   & 3 &  -1.0 $\pm$  & \,0.4 \cr
   & 4 &  6259  $\pm$ & \,335 \cr
\hline
\end{tabular}
\label{tab:vene}
\end{table*}

\clearpage

\renewcommand{\baselinestretch}{1}

\end{document}